\documentclass[fleqn,12pt]{article}
\usepackage{amssymb,latexsym}
\usepackage{amsmath, amsthm}
\usepackage[OT2,OT1]{fontenc}

\usepackage{url}
\usepackage{hyperref}
\input xy
\xyoption{all}
\DeclareFontFamily{OT1}{rsfs}{}
\DeclareFontShape{OT1}{rsfs}{m}{n}{ <-7> rsfs5 <7-10> rsfs7 <10->
rsfs10}{} \DeclareMathAlphabet{\mycal}{OT1}{rsfs}{m}{n}
\def\scri{{\mycal I}}%

\begin{document}
\title{\bf Asymptotic conformal Yano-Killing tensors for asymptotic anti-de Sitter
 spacetimes and conserved quantities}
\author{Jacek Jezierski\thanks{Partially supported by grant
 EPSRC: EP/D032091/1. E--mail: \texttt{Jacek.Jezierski@fuw.edu.pl}}
      }
\maketitle
{\catcode `\@=11 \global\let\AddToReset=\@addtoreset}
\AddToReset{equation}{section}
\renewcommand{\theequation}{\thesection.\arabic{equation}}

\newtheorem{Definition}{Definition}
\newtheorem{Lemma}{Lemma}
\newtheorem{Theorem}{Theorem}
\newtheorem{Remark}{Remark}
\newtheorem{Proposition}{Proposition}
\newcommand{\gn}[2]{\stackrel{\mbox{\tiny (#1)}}{#2}}
\newcommand{\gthree}{h}
\newcommand{\TBR}{T^{\scriptscriptstyle BR}}
\newcommand{\TEM}{T^{\scriptscriptstyle EM}}
\newcommand{\QBR}{CQ^{\scriptscriptstyle BR}}
\newcommand{\QEM}{CQ^{\scriptscriptstyle EM}}
\newcommand{\QYK}{\Theta}
\newcommand{\Img}{{\rm Image}}
\newcommand{\dtwo}{\mathbf{\Delta}}
\newcommand{\kolo}[1]{\vphantom{#1}\stackrel{\circ}{#1}\!\vphantom{#1}}
\newcommand{\eq}[1]{(\ref{#1})}
\newcommand{\arxiv}[1]{\url{http://arxiv.org/abs/#1}}
\newcommand{\rd}{\,{\rm d}} 
\newcommand{\tr}{\mathop {\rm tr}\nolimits }
\newcommand{\trg}{\mathop {\rm tr_g}\nolimits }
\newcommand{\arsinh}{\mathop {\rm arsinh}\nolimits }
\newcommand{\artanh}{\mathop {\rm artanh}\nolimits }
\newcommand{\be}{\begin{equation}}
\newcommand{\ee}{\end{equation}}
\newcommand{\ber}{\begin{eqnarray}}
\newcommand{\eer}{\end{eqnarray}}
\newcommand{\tg}{{\tilde g}}
\newcommand{\tGam}{{\tilde\Gamma}}
\newcommand{\tD}{{\tilde D}}
\newcommand{\tL}{{\tilde L}}
\newcommand{\N}{{\mathbb N}}
\newcommand{\tR}{{\tilde R}}
\newcommand{\mcY}{{\cal Y}}
\newcommand{\ped}{{\chi}}
\newcommand{\Exp}{{\rm e}}


\newcounter{mnotecount}[section]
\renewcommand{\themnotecount}{\thesection.\arabic{mnotecount}}
\newcommand{\mnote}[1]
{\protect{\stepcounter{mnotecount}}$^{\mbox{\footnotesize  $
      \bullet$\themnotecount}}$ \marginpar{\raggedright\tiny
    $\!\!\!\!\!\!\,\bullet$\themnotecount: #1} }
\newcommand{\JJ}[1]{\mnote{\textbf{JJ:} #1}}
\newcommand{\ptc}[1]{\mnote{\textbf{PTC:} #1}}

\begin{abstract}
Conformal rescaling of conformal Yano--Killing tensors and
relations between Yano and CYK tensors are discussed.
Pullback of these objects to a submanifold is used to construct all
solutions of a CYK equation in anti-de Sitter and de Sitter
spacetimes.
Properties of asymptotic conformal Yano--Killing tensors are
examined for asymptotic anti-de Sitter spacetimes. Explicit asymptotic forms
of them are derived. The results are used to construct asymptotic charges
in asymptotic AdS spacetime. Well known examples like Schwarzschild-AdS,
 Kerr-AdS and NUT-AdS are examined carefully in the construction of
 the concept of energy, angular momentum and dual mass in asymptotic AdS
spacetime.
\end{abstract}

\section{Introduction}
According to \cite{cykem} one can define, in terms of spacetime
curvature, two kinds of conserved quantities with the help of
conformal Yano--Killing tensors (see \cite{Tachibana}, \cite{Yano}).
Sometimes they are also called conformal
Killing forms or twistor forms (see e.g. \cite{BFGK}, \cite{Moroianu},
\cite{Semmelmann}, \cite{Stepanow}). The first kind is linear and
the second quadratic with respect to the Weyl tensor but a basis
for both of them is the Maxwell field. Conserved quantities which
are linear with respect to CYK tensor were investigated many times
(cf. \cite{Abbott-Deser}, 
\cite{Glass-Naber}, 
\cite{Goldberg1}, 
\cite{JJspin2}, \cite{kerrnut}, \cite{cykem}, \cite{Pen1}, 
\cite{Pen-Rin}). On the other hand, quadratic charges are less
known and 
usually examined in terms of the Bel--Robinson
tensor (see e.g. \cite{Berg}, \cite{Sen}, \cite{Ch-Kl},
\cite{Douglas}).

In electrodynamics the linear quantity corresponds to electric or
magnetic charge and the quadratic one expresses the energy, linear
momentum or angular momentum of the Maxwell field. In gravity both
kinds of charges play a role of energy. The linear conserved
quantities (as two-surface integrals) correspond to ADM mass and
linear or angular momentum but bilinear ones are not obviously
related to energy. They rather play a role of energy estimates
like in \cite{Ch-Kl} (cf. \cite{LarsAnd}). In this paper we
analyze the existence and the properties of the linear charges for
asymptotic anti-de Sitter spacetimes.

Let $M$ be an $n$-dimensional ($n>1$) manifold with a Riemannian
or pseudo-Riemannian metric $g_{\mu\nu}$. The covariant derivative
associated with the~Levi--Civita connection will be denoted by
$\nabla$ or just by ``$\,;\,$''. By $T_{...(\mu\nu)...}$ we will
denote the symmetric part and by $T_{...[\mu\nu]...}$ the
skew-symmetric part of tensor $T_{...\mu\nu...}$ with respect to
indices $\mu$ and $\nu$ (analogous symbols may be used for more
indices).

Let $Q_{\mu\nu}$ be a skew-symmetric tensor field (two-form) on
$M$ and let us denote by ${\cal Q}_{\lambda \kappa \sigma}$ a
(three-index) tensor which is defined as follows:
\begin{equation}\label{CYK_eq1}
    {\cal Q}_{\lambda \kappa
    \sigma}(Q,g):= Q_{\lambda \kappa ;\sigma} +Q_{\sigma \kappa
    ;\lambda} - \frac{2}{n-1} \left( g_{\sigma
    \lambda}Q^{\nu}{_{\kappa ;\nu}} + g_{\kappa (\lambda }
    Q_{\sigma)}{^{\mu}}{_{ ;\mu}} \right) \, .
\end{equation}
The object $\cal Q$ has the following algebraic properties
\be\label{wlQ}
   {\cal Q}_{\lambda\kappa\mu}g^{\lambda\mu}=0=
   {\cal Q}_{\lambda\kappa\mu}g^{\lambda\kappa} \, , \quad
{\cal Q}_{\lambda\kappa\mu} =
{\cal Q}_{\mu\kappa\lambda}\, ,
   \ee
i.e. it is traceless and partially symmetric.
\begin{Definition}\label{CYK_df}
    A skew-symmetric tensor $Q_{\mu\nu}$ is a conformal Yano--Killing tensor
    (or simply CYK tensor) for the metric $g$ iff
    \mbox{${\cal Q}_{\lambda \kappa \sigma}(Q,g) = 0$}.
\end{Definition}
In other words, $Q_{\mu\nu}$ is a conformal Yano--Killing tensor if
it fulfils the following equation:
\begin{equation}\label{CYK_eq2}
    Q_{\lambda \kappa ;\sigma} +Q_{\sigma \kappa ;\lambda} =
    \frac{2}{n-1} \left( g_{\sigma \lambda}Q^{\nu}{_{\kappa ;\nu}} +
    g_{\kappa (\lambda } Q_{\sigma)}{^{\mu}}{_{ ;\mu}} \right) \,
\end{equation}
(first proposed by Tachibana and Kashiwada, cf. \cite{Tachibana}).

A more abstract way {with no indices} of describing a CYK
tensor can be found in \cite{BCK}, \cite{Moroianu},
\cite{Semmelmann} or \cite{Stepanow},
where it is considered as the element of the kernel of the twistor operator $Q
\rightarrow {\cal T}_{wist} Q$ defined as follows:
\[ \forall X \;\; {\cal T}_{wist} Q (X) :=
\nabla_X Q -\frac1{p+1} X \lrcorner \rd Q +
\frac1{n-p+1}  g(X) \wedge \rd^* Q \, .\]
However, to simplify the exposition, we prefer abstract index notation
which also seems to be more popular.

The paper is organized as follows: In Section 2
we prove the Theorem:
if $Q$ is a CYK tensor of the ambient $(n+1)$-dimensional spacetime metric
$\gn{n+1}{g}$ (in a special form), then its pullback to (the correctly
chosen) submanifold 
is a CYK tensor of the induced metric $\gn{n}{g}$, which may be
easily applied for  anti-de Sitter embedded in flat
pseudo-Riemannian five-dimensional manifold. In the next Section
we use the preceding results to construct all CYK tensors in de
Sitter and anti-de Sitter spacetimes. Section 4 is devoted
to the anti-de Sitter spacetime together with some important
examples. In particular, we construct (more explicitly than usual)
Fefferman-Graham canonical coordinates for the Kerr-AdS solution.
Next Section contains analysis of symplectic structure at scri and,
finally, in the Section \ref{asch} we analyze asymptotic charges.
To clarify the exposition some of the technical results and proofs have been
shifted to the appendix.

\section{Pullback of CYK tensor to submanifold of codimension one}
Let $N$ be a differential manifold of dimension $n+1$ and
$\gn{n+1}{g}$ its metric tensor (the signature of the metric
plays no role). Moreover, we assume that there exists a coordinate system
$(x^A)$, where $A=0,\dots,n$, in which $\gn{n+1}{g}$ takes the following form:
\begin{equation}\label{higher_dim_metric}
    \gn{n+1}{g} = f(u)h + s\rd u^2,
\end{equation}
where $s$ is equal to 1 or $-1$, $u\equiv x^n$, $f$ is a certain
function, and $h$ is a certain tensor, which does not depend on
$u$. 
The metric (\ref{higher_dim_metric}) possesses
a conformal Killing vector field\footnote{A
conformally rescaled metric ${\tilde{g}} =\frac1f\gn{n+1}g = h + s\frac{\rd
u^2}{f(u)}= h + s\rd v^2$ (where $\rd v:=f(u)^{-1/2}\rd u$) has
the Killing vector $\partial_v=\sqrt{f}\partial_u$, which is a conformal Killing
vector field for
the original metric (\ref{higher_dim_metric}).}
$\sqrt{f}\partial_u$.
 Tensor $f(u)h$ is a metric tensor
on a submanifold $M:=\{ u =\textrm{const.}\}$. We will denote it by
$\gn{n}{g}$. We will distinguish all objects associated with the
metric $\gn{n}{g}$ by writing $(n)$ above their symbols. Similar
notation will be used for objects associated with the metric
$\gn{n+1}{g}$.

It turns out that:
\begin{Theorem}
If $Q$ is a CYK tensor of the metric
$\gn{n+1}{g}$ in $N$, then its pullback to the submanifold
 $M$ 
is a CYK tensor of the metric $\gn{n}{g}$.
\end{Theorem}
\begin{proof}
 In
order to show this, we need to derive some helpful formulae. Let
us notice that in coordinates $x^A$ we have\footnote{In this
chapter we will use the convention that indices denoted by capital
letters of the Latin alphabet go from $0$ to $n$ and Greek indices
go from $0$ to $n-1$. The index $u$ denotes $n$-th component of a tensor.}:
\begin{equation}\label{higher_dim_metric_compts}
    \gn{n+1}{g}\!\!\!{_{uu}} = s, \quad
    \gn{n+1}{g}\!\!\!{_{u\mu}}=0, \quad \gn{n+1}{g}\!\!\!{_{\mu\nu}}=\gn{n}
    {g}_{\mu\nu}\,.
\end{equation}
It means that the only non-vanishing derivatives of the metric are
the following:
\begin{equation}\label{higher_dim_metric_der}
    \quad \gn{n+1}{g}\!\!\!{_{\mu\nu,u}}=\Phi(u)\gn{n}{g}_{\mu\nu}
    \quad \textrm{and} \quad
    \quad \gn{n+1}{g}\!\!\!{_{\mu\nu,\rho}}=\gn{n}{g}_{\mu\nu,\rho}\,,
\end{equation}
where $\Phi(u)=\frac{\rd}{\rd u}\left(\log f(u)\right)$. Using the
formula
\[
    \gn{n+1}{\Gamma}\!{^A}{_{BC}}=\frac12 g^{AD}(g_{DB,C} + g_{DC,B} - g_{BC,D}),
\]
we compute all non-vanishing Christoffel symbol of the metric
$\gn{n+1}{g}$:
\begin{equation}\label{higher_dim_Christ}
    \gn{n+1}{\Gamma}\!{^A}{_{\nu u}} = \frac12\Phi(u)\delta^A{}_\nu\,, \quad
    \gn{n+1}{\Gamma}\!{^u}{_{\mu\nu}} = -\frac{s}{2}\Phi(u)g_{\mu\nu}, \quad
    \gn{n+1}{\Gamma}\!{^\mu}{_{\nu\rho}} = \gn{n}{\Gamma}{}^\mu{}_{\nu\rho}\,.
\end{equation}
Using formulae~(\ref{higher_dim_Christ}) we compute:
{\setlength\arraycolsep{2pt}
\begin{eqnarray}\label{higher_der_to lower}
    \gn{n+1}{\nabla}\!\!\!{_\mu} \, Q_{\nu\rho} & = & Q_{\nu\rho,\mu} - Q_{A\rho}\gn{n+1}
    {\Gamma}\!{^A}{}_{\nu\mu} - Q_{\nu A}\gn{n+1}{\Gamma}\!{^A}{}_{\rho\mu}\nonumber\\
    & = & \gn{n}{\nabla}_\mu Q_{\nu\rho} - Q_{u\rho}\gn{n+1}{\Gamma}\!{^u}{}_{\nu\mu}
    - Q_{\nu u}\gn{n+1}{\Gamma}\!{^u}{}_{\rho\mu}\nonumber\\
    & = & \gn{n}{\nabla}_\mu Q_{\nu\rho} + \frac{s}{2}\Phi(u)Q_{u\rho}g_{\mu\nu}
    + \frac{s}{2}\Phi(u) Q_{\nu u}g_{\rho\mu}\,.
\end{eqnarray}}
Let us denote $\gn{n}{\xi}_{\rho}:= g^{\mu\nu} \gn{n}{\nabla}_\mu Q_{\nu\rho}$.
The formula~(\ref{higher_der_to lower}) directly implies that:
\begin{equation}\label{higher_dim_div2}
    g^{\mu\nu}\gn{n+1}{\nabla}\!\!\!{_\mu} \, Q_{\nu\rho} = \gn{n}{\xi}_{\rho}
    + s\frac{n-1}{2}\Phi(u)Q_{u\rho}\,.
\end{equation}
Tensor $Q$ satisfies the CYK equation, i.e.
\begin{equation}\label{higher_dim_CYK_eq}
    \gn{n+1}{\nabla}\!\!\!{_A} \, Q_{BC} + \gn{n+1}{\nabla}\!\!\!{_B} \, Q_{AC}
    = \frac2{n}\left( g_{AB}\!\!\!\gn{n+1}{\xi}\!\!\!{_C}
    - g_{C(B}\!\!\!\gn{n+1}{\xi}\!\!\!{_{A)}}\right).
\end{equation}
Substituting $A=B=u$ and $C=\rho$ we get:
\begin{equation}\label{higher_dim_div3}
    \gn{n+1}{\nabla}\!\!\!{_u}\, Q_{u\rho} =
    \frac{s}{n}\gn{n+1}{\xi}\!\!\!{_\rho} \,.
\end{equation}
Using formulae~(\ref{higher_dim_div2})
and~(\ref{higher_dim_div3}) we compute:
{\setlength\arraycolsep{2pt}
\begin{eqnarray}\label{higher_dim_div4}
    \gn{n+1}{\xi}\!\!\!{_\rho} & = & g^{AC}\gn{n+1}{\nabla}\!\!\!{_C}\, Q_{A\rho} = g^{uu}
    \gn{n+1}{\nabla}\!\!\!{_u}\, Q_{u\rho} +
    g^{\mu\nu}\gn{n+1}{\nabla}\!\!\!{_\mu}\, Q_{\nu\rho}\nonumber\\
    & = & \frac1{n}\gn{n+1}{\xi}\!\!\!{_\rho} + \gn{n}{\xi}_{\rho} + s\frac{n-1}{2}\Phi(u)
    Q_{u\rho}\,
\end{eqnarray}}which implies:
\begin{equation}\label{higher_dim_div}
    \gn{n+1}{\xi}\!\!\!{_\rho} = \frac{n}{n-1}\gn{n}{\xi}_\rho
    + s\frac{n}{2}\Phi(u)Q_{u\rho}\,.
\end{equation}
Using formulae~(\ref{higher_der_to lower})
and~(\ref{higher_dim_div}) we get: {\setlength\arraycolsep{2pt}
\begin{eqnarray}\label{higher_lower_CYK}
    \gn{n+1}{\nabla}\!\!\!{_\sigma} \, Q_{\lambda\kappa} &\,
    + & \gn{n+1}{\nabla}\!\!\!{_\lambda}\,
    Q_{\sigma\kappa} - \frac2{n}\left( g_{\sigma\lambda}\!\!\!\gn{n+1}{\xi}\!\!\!{_\kappa}
    - g_{\kappa(\lambda}\!\!\!\gn{n+1}{\xi}\!\!\!{_{\sigma)}}\right) = {}\nonumber\\
    & = & \gn{n}{\nabla}\!{_\sigma}\, Q_{\lambda\kappa}
    + \gn{n}{\nabla}\!{_\lambda} \, Q_{\sigma\kappa}
    - \frac2{n-1}\left( g_{\sigma\lambda}\gn{n}{\xi}{_\kappa}
    - g_{\kappa(\lambda}\gn{n}{\xi}{_{\sigma)}}\right).
\end{eqnarray}}
Left-hand side of the equation~(\ref{higher_lower_CYK}) vanishes because
 $Q$ is a CYK tensor of the metric $\gn{n+1}{g}$. This implies that the
 right-hand side is also equal to zero, hence the pullback of $Q$
 to the surface $u=\textrm{const.}$
is a CYK tensor of the metric $\gn{n}{g}$.
\end{proof}

\section{CYK tensors in the de Sitter (and anti-de Sitter) spacetime}
In this Section we will discuss the problem of existence and basic
properties of CYK tensors for de Sitter and anti-de Sitter metrics.
These metrics are solutions to the vacuum Einstein equations with
the cosmological constant $\Lambda$ having a maximal symmetry
group. Therefore, they can be treated as a generalization of the flat
Minkowski metric to the case of nonzero cosmological constant. De
Sitter metric is a solution of Einstein equations with positive
cosmological constant. Anti-de Sitter metric corresponds to
negative cosmological constant. We will restrict ourselves to the
case of four dimensional metrics, although they can be defined for
manifolds of any dimension (cf.  \cite{HIM}, \cite{Skenderis}).
In dimension four we can express these
metrics with the use of coordinate system $(t,r,\theta,\phi)$ as follows
\begin{equation}\label{deSitter_metric}
    {\tilde g} =
    -\left(1-s\frac{r^2}{l^2}\right)\rd t^2
    + \frac{1}{1-s\frac{r^2}{l^2}} \rd r^2 + r^2
    \rd \Omega_2,
\end{equation}
where $\rd\Omega_2:=\rd\theta^2 + \sin^2\theta\rd\phi^2$ is a unit
sphere metric and $l$
is a certain constant related to the cosmological constant by the
formula $\Lambda=3s/l^2$. Moreover, $s$ is equal to 1 for de Sitter metric
and $-1$ for anti-de Sitter metric.
When $l$ goes to infinity,
the cosmological constant goes to zero and  the metric~(\ref{deSitter_metric})
approaches a flat metric, as one could expect.

Quite often it is more convenient to use another (rescaled) coordinate system
together with the following notation: $\bar{t}=\frac{t}{l}$,
$\bar{r}=\frac{r}{l}$. In coordinates
$(\bar{t},\bar{r},\theta,\phi)$ (anti-)de Sitter metric has the
following form:
\begin{equation}\label{deSitter_metric_bar}
    {\tilde g} = l^2\left[(-1+s\bar{r}^2)\rd \bar{t}^2 + \frac1{1-s\bar{r}^2}\rd \bar{r}^2
    +\bar{r}^2\rd\Omega_2\right].
\end{equation}

(Anti-)de Sitter manifold (denoted by $\tilde M$) is (by definition) Einstein
spacetime i.e. its Ricci tensor is proportional to the metric. Its Weyl
tensor vanishes which means it is a conformally flat metric (i.e.
there exists a conformal rescaling which brings it to the flat Minkowski
metric). As we have shown in \cite{JJML}, four-dimensional
Minkowski spacetime admits twenty-dimensional space of solutions of
the CYK equation. Moreover, the following
\begin{Theorem}\label{conf_resc_th}
    If $Q_{\mu\nu}$ is a CYK tensor for the metric $g_{\mu\nu}$,
    then $\Omega^3 Q_{\mu\nu}$ is a CYK tensor for
    the conformally rescaled metric
    $\Omega^2 g_{\mu\nu}$.
\end{Theorem} \noindent
(proved in \cite{JJML})
implies that (anti-)de Sitter also admits precisely twenty independent CYK
tensors\footnote{In general, this might be true only locally, e.g. in
\cite{InfeldSchild} one can find some global topological
difficulties with the conformally covariant solutions of Maxwell
and Dirac equations for cosmological models conformal to Minkowski
spacetime.}. We will show in the sequel
how to obtain them in an independent way and examine their basic
properties.

In order to do that we use the immersion of our
four-dimensional (anti-)de Sitter spacetime in five-dimensional
flat pseudo-Riemmannian manifold with signature $(s,1,1,1,-1)$. In order to make
formulae more legible we will use the following convention:
Greek indices $\mu, \nu, \ldots$
label spacetime coordinates in $\tilde M$ and run from 0 to 3;
Latin indices $i,j, \ldots$ label space coordinates and run
from 1 to 3, and finally indices denoted by capital letters of the Latin alphabet go
from $0$ to $4$ and they label coordinates in $N$.

Let $N$ be a five-dimensional differential manifold with a global
coordinate system $(y^A)$. We define the metric tensor $\eta$ of
the manifold $N$ by the formula:
\begin{equation}\label{deSitter_eta}
    \eta=\eta_{AB}\rd y^A \otimes \rd y^B=s \rd y^0 \otimes \rd y^0
    + \rd y^1\otimes \rd y^1 + \rd y^2\otimes \rd y^2
    + \rd y^3\otimes \rd y^3 - \rd y^4\otimes \rd y^4 \, .
\end{equation}
Let $\tilde M$ be a submanifold of $N$ defined by:
\begin{equation}\label{deSitter_submanifold}
    \eta_{AB}y^{A}y^{B}=s l^2.
\end{equation}
The metric $\eta$ restricted to $\tilde M$ is just the (anti-)de Sitter
metric (cf. \cite{Rindler}). In order to see this, let us introduce a coordinate
system $(\bar{t},\bar{r},\theta,\phi)$ on $\tilde M$. However, we need to
consider the cases $s=1$ and $s=-1$ separately. For $s=1$ a
parametrization of $\tilde M$ takes the following form:
\begin{equation}\label{deSitter_y0}
    y^0=l\sqrt{1-\bar{r}^2}\cosh\bar{t},
\end{equation}
\begin{equation}\label{deSitter_y1}
    y^1=l\bar{r}\sin\theta\cos\phi,
\end{equation}
\begin{equation}\label{deSitter_y2}
    y^2=l\bar{r}\sin\theta\sin\phi,
\end{equation}
\begin{equation}\label{deSitter_y3}
    y^3=l\bar{r}\cos\theta,
\end{equation}
\begin{equation}\label{deSitter_y4}
    y^4=l\sqrt{1-\bar{r}^2}\sinh\bar{t}.
\end{equation}
If $s=-1$, the analogous formulae are the following:
\begin{equation}\label{anti-deSitter_y0}
    y^0=l\sqrt{1+\bar{r}^2}\cos\bar{t},
\end{equation}
\begin{equation}\label{anti-deSitter_y1}
    y^1=l\bar{r}\sin\theta\cos\phi,
\end{equation}
\begin{equation}\label{anti-deSitter_y2}
    y^2=l\bar{r}\sin\theta\sin\phi,
\end{equation}
\begin{equation}\label{anti-deSitter_y3}
    y^3=l\bar{r}\cos\theta,
\end{equation}
\begin{equation}\label{anti-deSitter_y4}
    y^4=l\sqrt{1+\bar{r}^2}\sin\bar{t}.
\end{equation}
Let us notice that functions $l$, $\bar{t}$, $\bar{r}$, $\theta$
and $\phi$ can be considered as the local coordinate system on $N$. Substituting
formulae~(\ref{deSitter_y0})--(\ref{deSitter_y4})
or~(\ref{anti-deSitter_y0})--(\ref{anti-deSitter_y4}) into definition~(\ref{deSitter_eta})
of the metric $\eta$ we get:
\begin{equation}\label{deSitter_eta_bar}
    \eta =
    s\rd l^2 + l^2\left[(-1+s\bar{r}^2)\rd \bar{t}^2
    + \frac1{1-s\bar{r}^2}\rd {\bar r}^2
    +\bar{r}^2\rd\Omega_2\right]\, .
\end{equation}
In particular, formula~(\ref{deSitter_eta_bar}) implies that $\eta$
restricted to the surface $M:=\{ l=\textrm{const.} \}\subset N$ has the same form as
the metric $\tilde g$ (cf.~(\ref{deSitter_metric_bar})).

Identifying the (anti-)de Sitter spacetime with the submanifold
$\tilde M$ enables one to find all Killing vector fields of the metric $\tilde g$.
The vector fields
\[
    L_{AB} := y_A\frac{\partial}{\partial y^B} - y_B\frac{\partial}{\partial y^A}
\]
(where $y_A:=\eta_{AB} y^B$) are the Killing fields of the metric
$\eta$. However, the
formulae defining the fields $L_{AB}$ depend on the sign $s$.
For $s=1$ we get:
\begin{equation}\label{deSitter_killing40}
    L_{40}=-\frac{\partial}{\partial \bar{t}}\,,
\end{equation}
\begin{equation}\label{deSitter_killingi4}
    L_{i4}=\frac{x^i}{\sqrt{1-\bar{r}^2}}\cosh\bar{t}\frac{\partial}{\partial \bar{t}}
    + \sqrt{1-\bar{r}^2}\sinh\bar{t}\frac{\partial}{\partial x^i},
\end{equation}
\begin{equation}\label{deSitter_killingi0}
    L_{i0}=-\frac{x^i}{\sqrt{1-\bar{r}^2}}\sinh\bar{t}\frac{\partial}{\partial \bar{t}}
    - \sqrt{1-\bar{r}^2}\cosh\bar{t}\frac{\partial}{\partial x^i},
\end{equation}
\begin{equation}\label{deSitter_killingij}
    L_{ij}= x^i\frac{\partial}{\partial x^j} - x^j\frac{\partial}{\partial x^i},
\end{equation}
where in the coordinate system on $N$ instead of spherical
coordinates $\bar{r},\theta,\phi$
we use Cartesian $x^k:=\frac{y^k}{l}={\bar r}n^k$, $k=1,2,3$.

If $s=-1$ in coordinate system $(l,\bar{t}, x^k)$ we have:
\begin{equation}\label{anty-deSitter_killing40}
    L_{40}=\frac{\partial}{\partial \bar{t}}\,,
\end{equation}
\begin{equation}\label{anty-deSitter_killingi4}
    L_{i4}=  \frac{x^i}{\sqrt{1+\bar{r}^2}}\cos\bar{t}\frac{\partial}{\partial \bar{t}}
    + \sqrt{1+\bar{r}^2}\sin\bar{t}\frac{\partial}{\partial x^i},
\end{equation}
\begin{equation}\label{anty-deSitter_killingi0}
    L_{i0}=-\frac{x^i}{\sqrt{1+\bar{r}^2}}\sin\bar{t}\frac{\partial}{\partial \bar{t}}
    + \sqrt{1+\bar{r}^2}\cos\bar{t}\frac{\partial}{\partial x^i},
\end{equation}
\begin{equation}\label{anty-deSitter_killingij}
    L_{ij}= x^i\frac{\partial}{\partial x^j} - x^j\frac{\partial}{\partial x^i}.
\end{equation}
It is easy to notice that those fields are tangent to $\tilde M$ and
therefore their restrictions to the submanifold are Killing fields
of the induced metric. The fields defined on $N$ as well as their
restrictions to $\tilde M$ will be denoted by the same symbol $L_{AB}$.
Restricting the fields $L_{AB}$ to $\tilde M$ we get 10 linearly
independent Killing fields of the metric $\tilde g$. This is the maximum
number of the independent Killing fields the four-dimensional
metric can have, so $L_{AB}$ span the space of the Killing fields
of the metric $\tilde g$.

The formula~(\ref{deSitter_eta_bar}) shows that the metric $\eta$
in the coordinates $(l,\bar{t},\bar{r},\theta,\phi)$ has the
form~(\ref{higher_dim_metric}) which implies that the CYK tensors of the
metric $\eta$ restricted to the surface $l=\textrm{const.}$ are
the CYK tensors of the induced metric. In this way from the CYK tensors
in $N$ we obtain the CYK tensors in $\tilde M$. Let us consider 10
linearly independent tensors $\rd y^A \wedge\rd y^B$ defined in
$N$. Obviously they are Yano tensors of the metric $\eta$. Their
restriction to the submanifold $\tilde M$ gives us 10 linearly
independent CYK tensors of $\tilde g$. None of them is Yano tensor. We
have\footnote{Tensors restricted to $\tilde M$ will be denoted by the
same symbols as the tensors defined on $N$. Hence
 $y^A$ can be treated as functions on $\tilde M$ defined by the
formulae~(\ref{deSitter_y0})--(\ref{deSitter_y4})
or~(\ref{anti-deSitter_y0})--(\ref{anti-deSitter_y4}), where $l$ is a
 constant.}:
\begin{eqnarray}
    \xi \quad \textrm{for} & \rd y^0 \wedge \rd y^i & \; \textrm{equals}
    \quad s\frac3{l^2}L_{0i}, \label{dy0dyi_div}\\[8pt]
    \xi \quad \textrm{for} & \rd y^0 \wedge \rd y^4 & \;\textrm{equals}
    \quad -s\frac3{l^2}L_{04}, \label{dy0dy4_div}\\[8pt]
    \xi \quad \textrm{for} & \rd y^i \wedge \rd y^j & \;\textrm{equals}
    \quad -s\frac3{l^2}L_{ij}, \label{dyidyj_div}\\[8pt]
    \xi \quad \textrm{for} & \rd y^i \wedge \rd y^4 & \;\textrm{equals}
    \quad s\frac3{l^2}L_{i4} \label{dyidy4_div}
\end{eqnarray}
(remember that according to the previous notation $\xi$ for a CYK
tensor $Q$ is defined by the formula
$\xi^\nu:=Q^{\mu\nu}{}_{;\mu}$). It turns out that all the tensors
of the form $*(\rd y^A \wedge\rd y^B)$ (where $*$ denotes Hodge
duality related to the metric $\tilde g$) are Yano tensors. Tensors $\rd
y^A \wedge\rd y^B$ and $*(\rd y^A \wedge\rd y^B)$ are linearly
independent and there are twenty of them, therefore they span the
space of all solutions of the CYK equation for the
(anti-)de Sitter metric.

At the end, we consider the correspondence between CYK tensors in
Minkowski spacetime and the solutions of CYK equation in (anti-)de
Sitter spacetime -- tensors $\rd y^A \wedge\rd y^B$ and $*(\rd
y^A \wedge\rd y^B)$. To be more precise, we examine the behaviour of
the coefficients of the latter when we pass to the limit
$l\to\infty$ (as we know, in this limit the metric $\tilde g$ becomes the
flat Minkowski metric). There is, however, a crucial issue we have
to mention. Any CYK tensor can be multiplied by a constant, but on
$\tilde M$ the function $l$ is constant. Therefore in order to obtain
finite, non-zero limit we have to multiply each CYK tensor by a
proper power of $l$. Finally we get

\begin{eqnarray}
\lim_{l\to\infty}(\rd y^i \wedge \rd y^j) & = & ({\cal T}_i\wedge{\cal T}_j),
    \label{lim_dyidyj}\nonumber\\
    \lim_{l\to\infty}*(\rd y^i \wedge \rd y^j) & = & *({\cal T}_i\wedge{\cal T}_j),
    \label{lim_dyidyj_star}\nonumber\\
    \lim_{l\to\infty}(\rd y^i \wedge \rd y^4) & = & ({\cal T}_0\wedge{\cal T}_i),
    \label{lim_dyidy4}\nonumber\\
    \lim_{l\to\infty}*(\rd y^i \wedge \rd y^4) & = & *({\cal T}_0\wedge{\cal T}_i),
    \label{lim_dyidy4_star} \label{ytoT}\\
    \lim_{l\to\infty}(l \rd y^0 \wedge \rd y^i) & = & -s ({\cal D}\wedge{\cal T}_i),
    \label{lim_dy0dyi}\nonumber\\
    \lim_{l\to\infty}*(l \rd y^0 \wedge \rd y^i) & = & -s*({\cal D}\wedge{\cal T}_i),
    \label{lim_dy0dyi_star}\nonumber\\
    \lim_{l\to\infty}(l \rd y^0 \wedge \rd y^4) & = & s({\cal D}\wedge{\cal T}_0),
    \label{lim_dy0dy4}\nonumber\\
    \lim_{l\to\infty}*(l \rd y^0 \wedge \rd y^4) & = & s*({\cal D}\wedge{\cal T}_0),
    \label{lim_dy0dy4_star}\nonumber
\end{eqnarray}
where the space of Killing fields is spanned by the fields
\begin{equation}\label{kvf_generators}
{\cal T}_\mu := \frac{\partial}{\partial x^{\mu}},\quad
{\cal L}_{\mu\nu} := x_{\mu}\frac{\partial}{\partial x^{\nu}} -
x_{\nu}\frac{\partial}{\partial x^{\mu}}
\end{equation}
(here $(x^{\mu})$ are Cartesian coordinates,
 $x_{\mu}=\eta_{\mu\nu}x^{\nu}$, $\eta_{\mu\nu}:=
\textrm{diag}(-1,1,1,1)$) and
\begin{equation}\label{dil_generators}
{\cal D} := x^{\mu}\frac{\partial}{\partial x^{\mu}}\,
\end{equation}
is a {\em dilation vector field}.

\underline{Remark}: The formulae \eq{ytoT} imply that different CYK
tensors in the (anti-)de Sitter metric may converge to the same
tensor in Minkowski spacetime, e.g. $*(\rd y^1 \wedge \rd y^2)$
and $\rd y^4 \wedge \rd y^3$
 go to
$*({\cal T}_1\wedge{\cal T}_2)={\cal T}_3\wedge{\cal T}_0$,
 although $*(\rd y^1 \wedge \rd y^2)$
differs from $\rd y^4 \wedge \rd y^3$.

Moreover, we obtain the rest of CYK tensors in the Minkowski
spacetime as follows:

\begin{eqnarray}
\lim_{l\to\infty}l^2( \rd y^1 \wedge \rd y^2 - *\rd y^3 \wedge \rd y^4) & = &
-s \widetilde{\cal L}_{12}\, ,\label{l12}\\
\lim_{l\to\infty}l^2( \rd y^1 \wedge \rd y^3 - *\rd y^4 \wedge \rd y^2) & = &
-s \widetilde{\cal L}_{13}\, ,\nonumber\\
\lim_{l\to\infty}l^2( \rd y^2 \wedge \rd y^3 - *\rd y^1 \wedge \rd y^4) & = &
-s \widetilde{\cal L}_{23}\, ,\nonumber\\
\lim_{l\to\infty}l^2( \rd y^4 \wedge \rd y^1 - *\rd y^2 \wedge \rd y^3) & = &
s \widetilde{\cal L}_{01}\, ,\nonumber\\
\lim_{l\to\infty}l^2( \rd y^4 \wedge \rd y^2 - *\rd y^3 \wedge \rd y^1) & = &
s \widetilde{\cal L}_{02}\, ,\nonumber\\
\lim_{l\to\infty}l^2( \rd y^4 \wedge \rd y^3 - *\rd y^1 \wedge \rd y^2) & = &
s \widetilde{\cal L}_{03}\, ,\nonumber
\end{eqnarray}
where
\begin{equation}
\widetilde{\cal L}_{\mu\nu}:={\cal D}\wedge{\cal
L}_{\mu\nu}-\frac12 \eta({\cal D},{\cal D}){\cal T}_\mu\wedge{\cal
T}_\nu 
\end{equation}
(and $s=1$ for de Sitter, $s=-1$ for anti-de Sitter respectively).
The above formulae show how to obtain all CYK tensors in Minkowski spacetime
from the solutions of CYK equation in (anti-)de Sitter spacetime.

\section{Asymptotic anti-de Sitter spacetime}

For asymptotic analysis
let us change the radial coordinate in
the anti-de Sitter metric (\ref{deSitter_metric})
as follows
\[ z:= { l \over r+\sqrt{ {r^2}+{l^2} } } \, , \quad
 {\bar r}= \frac{r}l =\frac{1-z^2}{2z} \, , \]
which implies that
\begin{equation}\label{gAdS} {\tilde g}_{\mbox{\tiny \rm AdS}}=
\frac{l^2}{z^2} \left[ \rd z^2 -\left(\frac{1+z^2}{2}\right)^2\rd {\bar t}^2 +
 \left(\frac{1-z^2}{2}\right)^2 \rd\Omega_2 \right] \, .
 \end{equation}
The above particular form of ${\tilde g}_{\mbox{\tiny \rm AdS}}$ is well adopted to the
so-called conformal compactification
(see e.g. \cite{Hfriedrich}, \cite{Graham1}). More precisely,
the metric $g$ on the interior $\tilde M$ of a compact manifold $M$
with boundary $\partial M$
is said to be conformally compact if $g\equiv \Omega^2 {\tilde g}$
extends continuously (or with some degree of smoothness) as a metric to $M$,
where $\Omega$ is a defining function for the scri ${\scri}=\partial M$, i.e.
$\Omega >0$ on $\tilde M$ and $\Omega=0$, $\rd \Omega \neq 0$ on $\partial M$.
In the case of AdS metric (\ref{gAdS}) we have
\[ g_{\mbox{\tiny \rm AdS}}=\Omega^2 {\tilde g}_{\mbox{\tiny \rm AdS}},
 \quad \mbox{where} \quad \Omega:=\frac{z}l \, . \]

According to   \cite{FG}, \cite{Graham1} and \cite{Skenderis},
 our four-dimensional asymptotic AdS spacetime metric $\tilde g$
assumes in canonical
coordinates\footnote{Sometimes it is called Fefferman-Graham coordinate system.}
 the following form:
\be\label{canform} {\tilde g}={\tilde g}_{\mu\nu}\rd z^\mu \otimes\rd z^\nu
= \frac{l^2}{z^2}\left(
 \rd z\otimes\rd z + {\gthree}_{ab}\rd z^a \otimes \rd z^b \right) \ee
and the three-metric $h$ obeys the following asymptotic condition:
\begin{equation}\label{asc}
 {\gthree}={\gthree}_{ab}\rd z^a \otimes\rd z^b =
 {\stackrel{(0)}{h}} + z^2 {\stackrel{(2)}{h}}
 + z^3 \ped + O(z^4) \, .\end{equation}
 Let us observe that the term $\ped$ vanishes for the pure AdS given by (\ref{gAdS}).
 Moreover, the terms ${\stackrel{(0)}{h}}$ and ${\stackrel{(2)}{h}}$ have the
 standard form
\begin{eqnarray}
 {\stackrel{(0)}{h}} & = & \label{AdS1} \frac14 (\rd\Omega_2 -\rd {\bar t}^2)
 \, , \\  \label{AdS2}
  {\stackrel{(2)}{h}} & = & -\frac12 (\rd\Omega_2 +\rd {\bar t}^2) \, .
 \end{eqnarray}
 For generalized (asymptotically locally) anti-de Sitter spacetimes
 tensors ${\stackrel{(0)}{h}}$ and ${\stackrel{(2)}{h}}$ need not to be
 conformally ``trivial'', i.e. in the form
  (\ref{AdS1}) and (\ref{AdS2}) respectively. Such more general situation
    has been considered e.g.
 by Anderson, Chru{\'s}ciel \cite{genAdS}, Graham \cite{FG}, Skenderis
 \cite{Skenderis2}.
 Let us stress that in the general case only the induced metric
 ${\stackrel{(0)}{h}}$ may be changed freely beyond the conformal
 class, ${\stackrel{(2)}{h}}$ is always given by (\ref{2h}).
 Moreover, ${\stackrel{(0)}{h}}$ and $\chi$ form a symplectic structure
 on conformal boundary (cf. Section \ref{sf}). \\
 However, we assume the standard asymptotic AdS:
 The induced metric $h$ on $\scri$ is in the conformal class of the
 ``Einstein static universe'', i.e.
 \be\label{Esu}
  {\stackrel{(0)}{h}} = \exp(\omega)(\rd\Omega_2 -\rd {\bar t}^2)
  \ee
 for some smooth function $\omega$. This implies that our $\scri$ is
 a timelike boundary.

We use the following convention: Greek indices $\mu, \nu, \ldots$
label spacetime coordinates in $\tilde M$ and run from 0 to 3; Latin indices
$a,b, \ldots$ label coordinates on a tube $S:=\{ z=$ const.$\}$  and run from
0 to 2.

Functions $y^A$ given by equations (\ref{anti-deSitter_y0}--\ref{anti-deSitter_y4})
and restricted to $\tilde M$
can be expressed in coordinate system
$(z^\mu)\equiv (z^0,z^1,z^2,z^3)\equiv ({\bar t}, \theta, \phi, z)$
as follows
\begin{equation}\label{y0}
    y^0=\Omega^{-1} \frac{1+z^2}2\cos\bar{t}\, ,
\end{equation}
\begin{equation}\label{y1}
    y^k=\Omega^{-1} \frac{1-z^2}2 n^k\, ,
\end{equation}
\begin{equation}\label{y4}
    y^4=\Omega^{-1} \frac{1+z^2}2 \sin\bar{t}\, ,
\end{equation}
where $k=1,2,3$, and
\[ n:=\left[ \begin{array}{c}
 \sin\theta\cos\phi\\ \sin\theta\sin\phi \\ \cos\theta
 \end{array}\right]
 \]
is a radial unit normal in Euclidean three-space (identified with a point on
a unit sphere parameterized by coordinates $(\theta,\phi)$).

Let us denote  a CYK tensor in AdS spacetime by
${^{[AB]}}\tilde Q := l \rd y^A \wedge \rd y^B$.
Coordinates $y^A$ restricted to $\tilde M$, given by equations (\ref{y0}--\ref{y4}),
lead to the following explicit formulae for two-forms ${^{[AB]}}{\tilde Q}$:
\begin{eqnarray}\label{04}
 {}^{[04]}\tilde Q &=& \frac14 \Omega^{-3} (1-z^4)\rd {\bar t}\wedge\rd z \, ,\\
 \label{jk}
 {}^{[jk]}\tilde Q &=& \frac14 \Omega^{-3}
 \left[ (1-z^4)(n^j\rd n^k - n^k\rd n^j)\wedge\rd z +
 z(1-z^2)^2\rd n^j\wedge\rd n^k \right]  ,\\
 \label{0k}
 {}^{[0k]}\tilde Q &=& \frac14 \Omega^{-3} \left[
 (1-z^2)^2\cos{\bar t} \rd n^k \wedge \rd z +
 n^k (1+z^2)^2\sin{\bar t} \rd{\bar t} \wedge \rd z \right. \\ \nonumber
 & & +\left. z(1-z^4)\sin{\bar t}\rd n^k\wedge\rd {\bar t} \right]  ,\\
 \label{4k}
 {}^{[4k]}\tilde Q &=& \frac14 \Omega^{-3} \left[
 (1-z^2)^2\sin{\bar t} \rd n^k \wedge \rd z -
 n^k (1+z^2)^2\cos{\bar t} \rd{\bar t} \wedge \rd z \right. \\ \nonumber
 & & -\left. z(1-z^4)\cos{\bar t}\rd n^k\wedge\rd {\bar t} \right]  .
 \end{eqnarray}
 The other ten solutions we get applying Hodge star isomorphism.
 More precisely, an orthonormal frame
 $e^0:=\Omega^{-1}\left(\frac{1+z^2}2\right)\rd{\bar t}$,
 $e^1:=\Omega^{-1}\left(\frac{1-z^2}2\right)\rd \theta$,
 $e^2:=\Omega^{-1}\left(\frac{1-z^2}2\right)\sin\theta\rd\phi$,
 $e^3:=\Omega^{-1}\rd z$
 for the metric tensor (\ref{gAdS}), i.e.
 ${\tilde g}_{\mbox{\tiny \rm AdS}} = - e^0\otimes e^0 +
 \sum_{k=1}^3 e^k\otimes e^k$,
 enables one to calculate Hodge dual in a simple way, i.e.
 $*(e^0\wedge e^1)= - e^2\wedge e^3$, $*(e^0\wedge e^2)= - e^3\wedge e^1$,
 $*(e^0\wedge e^3)= - e^1\wedge e^2$, $*(e^1\wedge e^2)=  e^0\wedge e^3$,
 $*(e^2\wedge e^3)=  e^0\wedge e^1$, $*(e^3\wedge e^1)=  e^0\wedge e^2$.
 Moreover,
 \[ \rd n = \left[ \begin{array}{c} \cos\theta\cos\phi\rd\theta
 -\sin\theta\sin\phi\rd\phi \\
 \cos\theta\sin\phi\rd\theta + \sin\theta\cos\phi\rd\phi \\
 -\sin\theta\rd\theta
 \end{array}\right] = \frac{2\Omega}{1-z^2}\left[
 \begin{array}{c} \cos\theta\cos\phi \, e^1
 -\sin\phi \, e^2 \\
 \cos\theta\sin\phi \, e^1 + \cos\phi \, e^2 \\
 -\sin\theta \, e^1
 \end{array}\right] \, .
 \]
 Finally, for the dual two-forms $\ast\tilde Q$ we have
\begin{eqnarray}\label{*04}
 *{}^{[04]}\tilde Q &=& \left(\frac{1-z^2}{2\Omega}\right)^3
 \sin\theta\rd \theta \wedge\rd \phi \, , \\
 \label{*jk}
 *{}^{[jk]}\tilde Q &=&  \frac{1+z^2}{2\Omega^{3}}\rd {\bar t}
 \wedge \left[ z n^l\rd z - \frac{1-z^4}{4}\rd n^l \right] \epsilon_{jkl} \, , \\
 \label{*0k}
 *{}^{[0i]}\tilde Q &=&  \frac{1-z^2}{2\Omega^{3}} \left[
  \left(\frac{1-z^4}{4}\cos{\bar t}\rd{\bar t}+z\sin{\bar t}\rd z\right)
  \wedge n^j\rd n^k \right.\\ & &
  - \left. \frac{1-z^4}{8}\sin{\bar t}\rd n^j\wedge\rd n^k
  \right]\epsilon_{ijk} \, ,\nonumber \\
 \label{*4k}
 *{}^{[4i]}\tilde Q &=&  \frac{1-z^2}{2\Omega^{3}}
 \left[z \cos{\bar t}\, n^j\rd n^k\wedge\rd z
  -\sin {\bar t} \left(\frac{1-z^4}{4}\right)
  n^j\rd n^k\wedge\rd {\bar t} \right. \\ & & \left.
   + \cos{\bar t}
 \left(\frac{1-z^4}{8}\right)\rd n^j\wedge\rd n^k \right]\epsilon_{ijk} \, ,
 \nonumber
 \end{eqnarray}
where
\[
\epsilon_{ijk}:=\begin{cases}
+1 & \mbox{if}\; ijk \; \mbox{is an even permutation of}\; 1,2,3\\
-1 & \mbox{if}\; ijk \; \mbox{is an odd permutation of}\;  1,2,3\\
\,\ \ 0 & \mbox{in any other cases} \end{cases} \, .
\]
According to Theorem \ref{conf_resc_th}
 for conformally rescaled metric $g_{\mbox{\tiny\rm AdS}}$
 we get conformally related CYK tensors $Q:=\Omega^{-3}\tilde Q$.
 Their boundary values at conformal infinity ${\scri}:= \{ z=0 \}$
  take the following form:
 \begin{eqnarray}\label{Q04}
 {}^{[04]} Q \big|_{z=0} &=& \label{Qz04}\frac14 \rd {\bar t}\wedge\rd z
 \, ,\\
 \label{Qjk}
 {}^{[jk]}Q \big|_{z=0}&=& \frac14(n^j\rd n^k - n^k\rd n^j)\wedge\rd z
  \, ,\label{Qzjk}\\
 \label{Q0k}
 {}^{[0k]} Q \big|_{z=0}&=& \frac14 \left( \cos{\bar t} \rd n^k \wedge \rd z +
 n^k \sin{\bar t} \rd{\bar t} \wedge \rd z \right)
 \, ,\label{Qz0k}\\
 \label{Q4k}
 {}^{[4k]} Q \big|_{z=0}&=& \frac14  \left( \sin{\bar t} \rd n^k \wedge \rd z -
 n^k \cos{\bar t} \rd{\bar t} \wedge \rd z \right)
 \, .\label{Qz4k}
 \end{eqnarray}
 In Section \ref{asch}, when we define charges associated with CYK tensors,
 it will be clear that (\ref{Q04}) corresponds to the total energy and
 (\ref{Qjk}) to the angular momentum.
From this point of view CYK tensors
(\ref{Qz0k}-\ref{Qz4k}) correspond to the linear momentum and static moment.
Similarly, for dual conformally related CYK tensors
$\ast Q:=\Omega^{-3} \ast\tilde Q$ we obtain the following boundary values at
conformal infinity:
 \begin{eqnarray}\label{Q*04}
 *{}^{[04]} Q\big|_{z=0} &=& \frac{1}8
 \sin\theta\rd \theta \wedge\rd \phi \, ,\\
 \label{Q*jk}
 *{}^{[jk]} Q \big|_{z=0}&=&  \frac{1}8 \epsilon_{jki} \rd n^i
 \wedge \rd {\bar t}  \, ,\\
 \label{Q*0k}
 *{}^{[0i]} Q \big|_{z=0}&=&  \frac{1}8 \epsilon_{ijk}\left[
  \cos{\bar t}\rd{\bar t} \wedge n^j\rd n^k -
  \frac{1}{2}\sin{\bar t}\rd n^j\wedge\rd n^k
  \right]   ,\\
 \label{Q*4k}
 *{}^{[4i]} Q \big|_{z=0} &=&  \frac{1}8 \epsilon_{ijk}\left[
  \frac{1}{2}\cos{\bar t}\rd n^j\wedge\rd n^k -\sin {\bar t}\,
  n^j\rd n^k\wedge\rd {\bar t}  \right] .
 \end{eqnarray}
Let us notice that the ``rotated in time'' boundary values for
$*{}^{[0i]} Q$, $*{}^{[4i]} Q$
\begin{eqnarray}
 \left(  *{}^{[0i]} Q \cos{\bar t}
 + *{}^{[4i]} Q \sin{\bar t}\right) \big|_{z=0} &=&
 \frac{1}8 \epsilon_{ijk} \rd{\bar t} \wedge n^j\rd n^k \, ,
    \\
 \left(  *{}^{[4i]} Q \cos{\bar t} - *{}^{[0i]} Q \sin{\bar t}
 \right) \big|_{z=0} &=&  \frac{1}{16}
 \epsilon_{ijk}\rd n^j\wedge\rd n^k
 \end{eqnarray}
 and respectively for ${}^{[4i]} Q$, ${}^{[0i]} Q$
 \begin{eqnarray}
 \left(  {}^{[0k]} Q \cos{\bar t}
 + {}^{[4k]} Q \sin{\bar t}\right) \big|_{z=0} &=&
 \frac{1}4 \rd n^k\wedge\rd z \, ,
    \\
 \left(  {}^{[4k]} Q \cos{\bar t} - {}^{[0k]} Q \sin{\bar t}
 \right) \big|_{z=0} &=&  \frac{1}4  n^k \rd z\wedge \rd{\bar t}
 \end{eqnarray}
significantly simplify.

We denote by $(z^M)$ the coordinates on a unit sphere $(M=1,2, z^1=\theta,
z^2=\phi)$
and by $\gamma_{MN}$ the round metric on a unit sphere:
\[ \rd\Omega_2=\gamma_{MN}\rd z^M\rd z^N =\rd\theta^2 +\sin^2\theta \rd
\phi^2\, . \]
Let us also denote by $\varepsilon^{MN}$
a two-dimensional skew-symmetric tensor on $S^2$ such that
$\sin\theta\varepsilon^{\theta\phi}=1$.
Boundary values for Killing vector fields $L_{AB}$ at $\scri$ are:
\begin{eqnarray}\label{L04}
 L_{40} \big|_{z=0} &=& \frac{\partial}{\partial {\bar t}} \, , \\
 \label{Ljk}
 L_{jk}  \big|_{z=0}&=& \epsilon_{jkl} \varepsilon^{NM} n^l_{,M}
 \frac{\partial}{\partial z^N} , \quad
 L_{12} \big|_{z=0} = \frac{\partial}{\partial \phi} \, ,\\
 \label{L0k}
 L_{i0}  \big|_{z=0}&=&
  \cos{\bar t}\gamma^{-1}(\rd n^i) -
  \sin{\bar t} n^i \frac{\partial}{\partial {\bar t}}
   \, , \\
 \label{L4k}
 L_{i4}  \big|_{z=0} &=& \sin{\bar t}\gamma^{-1}(\rd n^i) +
  \cos{\bar t} n^i \frac{\partial}{\partial {\bar t}} \, .
 \end{eqnarray} 
 Together with (\ref{Qz04}-\ref{Qz4k}) and (\ref{AdS1}) they lead to
 the following universal formula:
 \be\label{asQ} {}^{[AB]} Q = {\stackrel{(0)}{h}}(L^{AB})\wedge\rd z \, ,\ee
 where $L^{AB}:=\eta^{AC}\eta^{BD}L_{CD}$.
 Similarly,
 \be \label{asQ*} \ast{}^{[AB]} Q =
 L^{AB} \rfloor {\rm vol}({\stackrel{(0)}{h}}) \, ,
  \ee
 where ${\rm vol}({\stackrel{(0)}{h}}):= \sqrt{-\det {\stackrel{(0)}{h}}}
 \rd{\bar t}\wedge\rd\theta\wedge\rd\phi$ is
 a canonical volume three-form on $\scri$.

 Moreover,
 \begin{eqnarray} *{}^{[34]}\tilde Q &=& \nonumber \Omega^{-3}
 \left[ \sin {\bar t} \sin^2\theta \left(\frac{1+z^2}2\right)
 \left(\frac{1-z^2}2\right)^2\rd\phi \wedge \rd {\bar t} \right. \\ & &
 - \cos{\bar t} \cos\theta \sin\theta\left(\frac{1+z^2}2\right)
 \left(\frac{1-z^2}2\right)^2\rd\theta\wedge\rd\phi\\ & & \left.
  -z \cos{\bar t}\sin^2\theta\left(\frac{1-z^2}2\right)
  \rd\phi\wedge\rd z \right] \nonumber \\
  &=& \nonumber \Omega^{-3}
 \left[ \sin {\bar t} \left(\frac{1+z^2}2\right)\left(\frac{1-z^2}2\right)^2
  (n^1\rd n^2- n^2\rd n^1)\wedge\rd {\bar t} \right. \\ & &
 - \cos{\bar t} \left(\frac{1+z^2}2\right)
 \left(\frac{1-z^2}2\right)^2\rd n^1\wedge\rd n^2\\ & & \left.
  -z \cos{\bar t}\left(\frac{1-z^2}2\right)
  (n^1\rd n^2- n^2\rd n^1)\wedge\rd z \right] \nonumber
  \end{eqnarray}
and
\begin{eqnarray} {}^{[12]}\tilde Q
  &=& \nonumber \Omega^{-3}
 \left[ z\left(\frac{1-z^2}2\right)^2 \rd n^1\wedge\rd n^2 \right. \\
  & & \phantom{XXX} +
  \left. \left(\frac{1-z^2}2\right)\left(\frac{1+z^2}2\right)
  (n^1\rd n^2- n^2\rd n^1)\wedge\rd z \right] \, .
  \end{eqnarray}

Formula (\ref{l12}) suggests that CYK tensor $ {}^{[12]} Q - *{}^{[34]} Q$
should correspond to $\widetilde{\cal L}_{12}$ in Minkowski spacetime
hence the third ($z$-th) component of angular momentum may correspond to
$ {}^{[12]} Q - *{}^{[34]} Q$ instead of $ {}^{[12]} Q$
(see Section \ref{asch}).

\subsection{Examples}
Schwarzschild-AdS solution (cf. \cite{HM}):
\be\label{SAdS} ds^2= -\left(\frac{r^2}{l^2} +1 -\frac{2m}{r}\right) \rd t^2
+\left(\frac{r^2}{l^2} +1 -\frac{2m}{r}\right)^{-1}\rd r^2 + r^2\rd\Omega_2 \ee
may be transformed into the canonical form (\ref{canform})
with the help of the coordinate $z$ defined by the following elliptic integral:
\[ z=\exp\left( \int \frac{\rd w}{w\sqrt{1+w^2-bw^3}} \right) \, ,\]
where $b:=\frac{2m}l$ and $w:=\frac{l}r$.
 For the function $F$ implicitly defined by the following
conditions:
\[ F(b,0)=0 \, , \quad F(0,w)=-\frac{w}{1+w^2} + \arsinh w \, ,\]
\[ \frac{\partial F}{\partial w} = \frac{w^2}{(\sqrt{1+w^2}+\sqrt{1+w^2-bw^3})
\sqrt{1+w^2}\sqrt{1+w^2-bw^3}} \]
we have
\be\label{zw} z = \frac{w}{1+\sqrt{1+w^2}} \exp[b F(b,w)]\, . \ee
Let us change a temporal coordinate in $M$ to $\bar t=\frac{t}l$.
On surface $S$ the three-metric $h$ can be expressed as follows:
\[ h = \left( \frac{\exp(bF)}{1+\sqrt{1+w^2}}\right)^2
\left[  \rd\Omega_2 -(1+w^2-bw^3) \rd {\bar t}^2 \right] \]
 with the components given only in an implicit
 form\footnote{In order to have it explicitly
we should express variable $w$ in terms of $z$, i.e. we have to find the inverse
function $w(z)$ for $z(w)$ given by (\ref{zw}).}.
Moreover, the asymptotics of $F$:
\[ F(b,w) = \frac16 w^3 -\frac3{20} w^5 + \frac{b}{16}w^6 + O(w^7) \]
enable one to derive the asymptotic form (\ref{asc}) for the
three-metric ${\gthree}_{ab}$ in the Schwarzschild-AdS spacetime.
More precisely,
\[ h = \Exp^{2bF} \left\{ \left(\frac{1-z^2\Exp^{-2bF}}{2}\right)^2 \rd\Omega_2
-\left[   \left(\frac{1+z^2\Exp^{-2bF}}{2}\right)^2 -
  \frac{2b z^3\Exp^{-3bF}}{1-z^2\Exp^{-2bF}} \right] \rd {\bar t}^2 \right\} \]
and
\[ F=\frac43 z^3 \left( 1+O(z^2) \right) \]
give
\begin{eqnarray}
 {\stackrel{(0)}{h}} & = &
 \frac14 (\rd\Omega_2 -\rd {\bar t}^2) \, , \label{h0S} \\
  {\stackrel{(2)}{h}} & = &  -\frac12 (\rd\Omega_2 +\rd {\bar t}^2) \, ,
  \label{Sch-AdS2} \\
 \ped & = & \label{chiS} \frac{4m}{3l}(\rd\Omega_2 +2 \rd {\bar t}^2) \, .
 \end{eqnarray}


The solution of Einstein equations with mass, angular momentum
and negative cosmological constant is explicitly given by
\begin{equation}\label{KerrAdSg}
{\tilde g}_{\mbox{\tiny\rm Kerr-AdS}} =
{\tilde g}_{tt} \rd t^2 + 2{\tilde g}_{t\phi} \rd t \rd \phi + {\tilde g}_{rr} \rd r^2
+ {\tilde g}_{\theta\theta} \rd \theta^2 + {\tilde g}_{\phi\phi} \rd \phi^2\, ,
\end{equation}
where
\[
{\tilde g}_{tt} = -1 +{2mr \over \rho ^2}-\frac{r^2+a^2\sin^2\theta}{l^2},
\quad {\tilde g}_{t\phi} = -a\sin ^2 \theta \left(\frac{2mr}{\rho ^2}
-\frac{r^2+a^2}{l^2}\right) \, , \]
 \[ {\tilde g}_{rr} = {\rho ^2 \over \triangle +(r^2+a^2)\frac{r^2}{l^2}}, \quad
{\tilde g}_{\theta \theta} = \frac{\rho^2}{1-\frac{a^2\cos^2\theta}{l^2}},
\]
\begin{equation}\label{KerrAdScompts}
{\tilde g}_{\phi\phi} =
\sin^2\theta\left[(r^2+a^2)\left(1-\frac{a^2}{l^2}\right)
+{2mra^2\sin^2\theta\over\rho^2}\right],
\end{equation}
with $\rho ^2$ and $\triangle$ defined as follows:
\begin{equation}\label{Kerr_metric_symbols}
 \rho ^2 := r^2 + a^2 \cos ^2 \theta \quad \textrm{and} \quad
 \triangle := r^2 -2mr +a^2\, .
\end{equation}
Asymptotic behaviour of ${\gthree}_{ab}$ for Kerr-AdS is analyzed in Appendix A
and gives the following result:
\begin{eqnarray} \label{Kerr-AdS1}
{\stackrel{(0)}{h}}& = & \frac14\left(
{\bar a}\sin\bar\theta+{\sqrt{1-{\bar a}^2\cos^2\bar\theta}}\right)^2
\Big[ \frac{1}{1-{\bar a}^2\cos^2\bar\theta}{\rd\bar\theta^2} + 
\\ & & + \nonumber
2{\bar a}\sin^2\bar\theta \rd{\bar t}\rd\phi 
+\sin^2\bar\theta (1-{\bar a}^2) \rd\phi^2 - \rd{\bar t}^2 \Big] \, ,\\
 {\stackrel{(0)}{h}}& = & \label{cfh0}
\frac{1-{\bar a}^2}{4\left(1-{\bar a}\sin\Theta\right)^2}
\Big[ {\rd\Theta^2} +\sin^2\Theta \rd\Phi^2 - \rd{\bar t}^2 \Big] \, , \\
\ped & = & \nonumber  \frac{b\omega}3 \Big\{ 2\rd{\bar t}^2
  -4 {\bar a}\sin^2\bar\theta
 \rd\bar t \rd\phi + \left(1-{\bar a}^2+{3{\bar a}^2\sin^2\bar\theta}
   \right) \sin^2\bar\theta \rd\phi^2 \\
 & & + \frac{\rd\bar\theta^2}{1-{\bar a}^2\cos^2\bar\theta} \Big\} \, ,
 \end{eqnarray}
where the canonical coordinate systems $({\bar t}, {\bar\theta}, \phi, z)$
is precisely defined in Appendix A. Moreover, another system
  of coordinates $({\bar t},\Theta,\Phi,R)$ given by (\ref{APhi}-\ref{AR})
  enables one to check explicitly conformal flatness of ${\stackrel{(0)}{h}}$
  (see formula (\ref{cfh0}) above).
  Parameters $\bar a$, $b$ are the rescaled constants $a$ and $m$ respectively,
  ${\bar a}:=\frac{a}l$, $b:=\frac{2m}l$.

According to \cite{GP},
the Pleba\'nski--Demia\'nski metric for the case of black-hole spacetimes becomes
  \begin{equation}
  \begin{array}{l}
{\displaystyle \rd s^2={1\over\Omega^2}\left\{
{Q\over\rho^2}\left[\rd t- \left(a\sin^2\theta
+4{\bf l}\sin^2{\textstyle{\theta\over2}} \right)\rd\phi \right]^2
   -{\rho^2\over Q}\,\rd r^2 \right.
} \\[8pt]
  \hskip8pc {\displaystyle
 \left. -{\tilde P\over\rho^2} \Big[ a\rd t
  -\Big(r^2+(a+{\bf l})^2\Big)\rd\phi \Big]^2
-{\rho^2\over\tilde P}\sin^2\theta\,\rd\theta^2 \right\}, }
\end{array}
  \label{newMetric}
  \end{equation}
  where
  \begin{equation}
  \begin{array}{l}
  {\displaystyle \Omega=1-{\alpha\over\omega}({\bf l}+a\cos\theta)\,r } \, ,\\
  \rho^2 =r^2+({\bf l}+a\cos\theta)^2 \, ,\\
  \tilde P= \sin^2\theta\,(1-a_3\cos\theta-a_4\cos^2\theta) \, ,\\
  Q= {\displaystyle (\omega^2k+e^2+g^2) -2mr +\epsilon r^2
-2\alpha{n\over\omega} r^3 -\left(\alpha^2k+{\Lambda\over3}\right)r^4}
  \end{array}
  \label{newMetricFns}
  \end{equation}
  and
 \begin{equation}
 \begin{array}{l}
  {\displaystyle a_3= 2\alpha{a\over\omega}m -4\alpha^2{a{\bf l}\over\omega^2}
  (\omega^2k+e^2+g^2) -4{\Lambda\over3}a{\bf l} } \, ,\\
  {\displaystyle a_4= -\alpha^2{a^2\over\omega^2}(\omega^2k+e^2+g^2)
  -{\Lambda\over3}a^2 }
  \end{array}
 \label{a34}
 \end{equation}
 with $\epsilon$, $n$ and $k$ given by
 the following formulae:

\begin{eqnarray}
 &&\epsilon= {\omega^2k\over a^2-{\bf l}^2}+4\alpha{{\bf l}\over\omega}\,m
 -(a^2+3{\bf l}^2) \left[ {\alpha^2\over\omega^2}(\omega^2k+e^2+g^2)+{\Lambda\over3} \right],
 \label{epsilon} \\[8pt]
 &&n= {\omega^2k\,{\bf l}\over a^2-{\bf l}^2} -\alpha{(a^2-{\bf l}^2)\over\omega}\,m
 +(a^2-{\bf l}^2){\bf l} \left[ {\alpha^2\over\omega^2}\,(\omega^2k+e^2+g^2)
 +{\Lambda\over3} \right] \, ,
  \label{n}
 \end{eqnarray}
\begin{equation}
  \left( {\omega^2\over a^2-{\bf l}^2}+3\alpha^2{\bf l}^2 \right)\,k
  = 1 +2\alpha{{\bf l}\over\omega}\,m
  -3\alpha^2{{\bf l}^2\over\omega^2}(e^2+g^2) -{\bf l}^2\Lambda \, .
  \label{k}
  \end{equation}
 It is also assumed that $|a_3|$ and $|a_4|$ are sufficiently small that $\tilde P$
 has no additional roots with $\theta\in[0,\pi]$.
This solution contains eight arbitrary parameters $m$, $e$, $g$, $a$, ${\bf l}$, $\alpha$,
$\Lambda$ and $\omega$. Of these, the first seven can be varied independently, and
$\omega$ can be set to any convenient value if $a$ or ${\bf l}$ are not both zero.

When $\alpha=0$, (\ref{k}) becomes \ $\omega^2k=(1-{\bf l}^2\Lambda)(a^2-{\bf l}^2)$ \ and
hence (\ref{epsilon}) and (\ref{n}) become
 $$ \epsilon=1-({\textstyle{1\over3}}a^2+2{\bf l}^2)\Lambda \,,
 \qquad\qquad n={\bf l}+{\textstyle{1\over3}}(a^2-4{\bf l}^2){\bf l}\Lambda. $$
 The metric is then given by (\ref{newMetric}) with
  $$ \begin{array}{l}
  \Omega=1 \, ,\\
  \rho^2 =r^2+({\bf l}+a\cos\theta)^2 \, ,\\
  \tilde P=\sin^2\theta(1+{\textstyle{4\over3}}\Lambda a{\bf l}\cos\theta
  +{\textstyle{1\over3}}\Lambda a^2\cos^2\theta) \, ,\\
  Q= (a^2-{\bf l}^2+e^2+g^2) -2mr + r^2
  -\Lambda\Big[(a^2-{\bf l}^2){\bf l}^2+({1\over3}a^2+2{\bf l}^2)r^2+{1\over3}r^4\Big] .
  \end{array} $$
 This is exactly the Kerr--Newman--NUT--de~Sitter solution in the form which is
 regular on the half-axis $\theta=0$. It represents a non-accelerating black hole
 with mass~$m$, electric and magnetic charges $e$ and $g$,
 a rotation parameter $a$
 and a NUT parameter~${\bf l}$ in a de~Sitter or anti-de~Sitter background.
 It reduces to known forms when ${\bf l}=0$ or $a=0$ or $\Lambda=0$.

It would be nice to understand in what sense this solution is
asymptotically AdS when we choose $\Lambda$ negative.
We start the analysis of this question with a simplest nontrivial extension
of Schwarzschild-AdS, namely we assume the following form
of the metric:
\be\label{NUTAdS} ds^2 = \frac{l^2}{w^2}(1+{\bar l}^2 w^2)
\left[ \rd\Omega_2 + A^{-1}\rd w^2 - \frac{A}{(1+{\bar l}^2 w^2)^2}
\left(\rd {\bar t}-4{\bar l}\sin^2\frac\theta{2}\rd\phi\right)^2 \right] \ee
which corresponds to the choice $\alpha=a=e=g=0$, $\Lambda=-\frac3{l^2}$
 in (\ref{newMetric}). Moreover,
 \[ A:= 1+w^2(1+6{\bar l}^2) - bw^3 - (1+3{\bar l}^2){\bar l}^2 w^4 \, ,
 \quad {\bar l}:=\frac{{\bf l}}l \, , \]
and $b$, $w$, ${\bar t}$ were defined previously ($b:=\frac{2m}l$,
$w:=\frac{l}r$,  ${\bar t}:=\frac{t}l$).
The canonical coordinate $z$ is defined by the following integral:
\[ z=\exp\left( \int \frac{\rd w}{w}
\sqrt{\frac{1+{\bar l}^2 w^2}{A(w)}} \right) \]
which should be more deeply analyzed if we want to obtain
 $\chi$.
For ${\stackrel{(0)}{h}}$ the situation is much simpler
because $\scri$ corresponds to $z=w=0$ and
the induced metric at $\scri$ takes the following form:
\begin{equation}
 {\stackrel{(0)}{h}}  = \frac14 \left[\rd\Omega_2 -
 \left(\rd {\bar t}-4{\bar l}\sin^2\frac\theta{2}\rd\phi\right)^2 \right]
 \,  \label{h0NUT}
 \end{equation}
and its inverse
 \begin{equation}
 {\stackrel{(0)}{h}}{^{ab}}\partial_a\partial_b  = 4 \left[
 -\partial^2_{\bar t}+ \partial^2_\theta +\frac1{\sin^2\theta}
 \left( \partial_\phi+4{\bar l}\sin^2\frac\theta{2}\partial_{\bar t}\right)^2
 \right]\, . \label{hinvNUT}
 \end{equation}
All the above calculations in this Section bring us closer
(with the help of ``succesive approximation method'') to the
answer what is the asymptotics of ${\gthree}_{ab}$ in Kerr--NUT--AdS
spacetime. We know already the full asymptotics of Kerr-AdS, i.e. when
$l=0$ (see Appendix A), and we derive $\stackrel{(0)}{h}$,
$\stackrel{(2)}{h}$ for NUT-AdS, i.e. when $a=0$ (cf. (\ref{h0NUT}),
(\ref{2h}) and Appendix B).
This is enough to perform the analysis of asymptotic charges (see
Section \ref{asch}). We hope to extend our calculations for the full
Kerr--Newman--NUT--AdS spacetime in the future.

\subsection{Conformal rescaling}\label{ads1}

The  method exploits the properties of the asymptotic anti-de
Sitter spacetime under conformal transformations. Let us consider a
metric $\tg$ related to $g$ by a conformal rescaling:
\be\label{conf4} \tg_{\mu\nu} = \Omega^{-2} g_{\mu\nu} \quad
\Longleftrightarrow \quad \tg^{\mu\nu} = \Omega^{2} g^{\mu\nu}
\;.\ee It is straightforward to derive 
\begin{eqnarray} \tGam^\mu {}_{\nu\kappa} & = &
\Gamma ^\mu {}_{\nu\kappa} +
\delta^\mu {_\kappa} \partial_\nu U  +  \delta^\mu{_\nu}
\partial_\kappa U  - g_{\nu\kappa} \nabla^\mu  U  \, ,
\label{conf6} \end{eqnarray} where $\nabla$ denotes the covariant
derivative of $g$ and $ U := -\log\Omega$. Moreover, in the case of
asymptotic AdS we can choose $$\Omega:=\frac{z}l \; .$$

Riemann tensor is defined as usual:
\[\tilde R^\mu_{\;\;\nu\alpha\beta} = - \tilde\Gamma^\mu_{\;\;\nu\alpha,\beta}
+ \tilde\Gamma^\mu_{\;\;\nu\beta,\alpha} +
\tilde\Gamma^\mu_{\;\;\gamma\alpha}\tilde\Gamma^\gamma_{\;\;\beta\nu}
-\tilde\Gamma^\mu_{\;\;\gamma\beta}\tilde\Gamma^\gamma_{\;\;\alpha\nu}\;.\]

To analyse Einstein equations we shall use the following formula
(cf. chapter 3.7 in \cite{SKM}):
\be\label{CT} \tilde R_{\mu\nu} = R_{\mu\nu} -  U
^{;\alpha}{_{;\alpha}}g_{\mu\nu} + (n-2)\left(
 U _{;\mu} U _{;\nu} -    U _{;\gamma} U ^{;\gamma} g_{\mu\nu}
-  U _{;\mu \nu}\right)\;. \ee

A $(3+1)$-decomposition of the rescaled metric $g$ in canonical coordinates
takes a simple form:
\[ g = g_{\mu\nu}\rd z^\mu \otimes \rd z^\nu =
  {\gthree}_{ab}\rd z^a \otimes \rd z^b + \rd z \otimes \rd z \, , \]
 where $h$ is the induced metric on a tube $S$.

The extrinsic curvature
\[ K_{ab}:=\frac12 \partial_3 g_{ab} \]
of the surface $S$ 
enables one to derive the following Christoffel symbols for the
Levi-Civita connection of the metric $g_{\mu\nu}$:
\[ \Gamma^3_{ab}=-K_{ab} \; , \quad
\Gamma^a_{3b}=  K_{cb}{\gthree}{^{ca}} \, .\] Moreover,
the rest of them are as follows:
\[ \Gamma^3_{3\mu}=\Gamma^\mu_{33}=0 \; , \quad
\Gamma^a_{bc}=\Gamma^a_{bc}({\gthree}) \, . \]

Ricci tensor of the four-metric $g$ expresses in terms of initial data $(h,K)$
on $S$ as follows:
\be\label{R33} R_{33}= -\partial_3 K^a{_a}
-K^a{_b}K^b{_a} \, , \ee
\be \label{Rab} R_{ab}=
{\cal R}_{ab}(h) -\partial_3 K_{ab} + 2 K_{bc}K^c{_a}
    - K^c{_c} K_{ab} \, ,\ee
\be \label{R3a} R_{3a}= K^b{_{a|b}} - K^b{_{b|a}} \, , \ee
where by ``$|$'' we denote a covariant derivative with respect to the three-metric $h$
and ${\cal R}_{ab}(h)$ is its Ricci tensor.

Riemann tensor (curvature of the metric $g$):
\be\label{tt}
R^a{_{bcd}}= {\cal R}^a{_{bcd}}(h) + K_{bc} K^a{_d} - K_{bd} K^a{_c} \, ,
\ee
\be\label{nt}
R^3{_{abc}}=  K_{ab|c} - K_{ac|b} \, ,
\ee
\be\label{nn}
R^3{_{a3b}}= -\partial_3 K_{ab} + K_{ac} K^c{_b} \, .
\ee
In particular,
\[ R^0{_{303}}= -\partial_3 K^0{_0} - K^0{_a} K^a{_0} \, .\]

The conformal Weyl tensor is defined as follows (cf. \cite{HfriedrichC})
\begin{equation}
 W^\mu{_{\nu\lambda\rho}}= R^\mu{_{\nu\lambda\rho}} +
 g_{\nu[\lambda}S_{\rho]}{^\mu} - \delta^\mu{_{[\lambda}}S_{\rho]\nu} \, ,
 \end{equation}
 where
 \[ S_{\mu\nu}=R_{\mu\nu} -\frac16 R g_{\mu\nu} \, . \]

Gauss-Codazzi-Ricci equations (\ref{tt} -- \ref{nn}) imply:
\[ R_{acdb}h^{cd} = -{\cal R}_{ab}(h) + K^c{_c} K_{ab} - K_{bc}K^c{_a} \, , \]
\[ \varepsilon_a{^{cd}}R_{3bcd}= \varepsilon_a{^{cd}}
\left( K_{bd|c}-K_{bc|d}\right)=
   - 2 \varepsilon_a{^{cd}} K_{bc|d} \, ,\]
where $\varepsilon^{abc}$ is the Levi-Civita antisymmetric
tensor for the three-metric $h$ such that $\sqrt{-\det h}\,
\varepsilon^{012}=1$.

Components of the Weyl tensor (with respect to initial data on $S$)
one can nicely describe in terms of traceless symmetric tensors $E$ and $B$,
electric and magnetic parts of $W$ respectively.
The electric part can be derived as follows:
\begin{eqnarray}\label{WE}
 E_{ab}:= W_{a3b3}& = & W_{acdb}h^{cd} =  R_{acdb}h^{cd}+
 \left( h_{c[d}S_{b]}{^a} - h{_{a[d}}S_{b]c}\right) h^{cd} \\
 & = & -{\cal R}_{ab}(h) + K^c{_c} K_{ab} - K_{bc}K^c{_a}+ \frac12 S_{ab}+
 \frac12 h_{ab} S_{cd}h^{cd} \, ,
 \end{eqnarray}
 where
 \begin{eqnarray}
 S_{ab}&=&{\cal R}_{ab}(h) -\frac16 {\cal R}(h) h_{ab} + \frac16 h_{ab}
 \left( K^2 + 2\partial_3 K \right) \\ \nonumber
  & & + 2 K_{bc}K^c{_a} - \partial_3 K_{ab} - K K_{ab}
  \end{eqnarray}
 as a consequence of (\ref{R33}-\ref{R3a}) and $K:= K^c{_c}$.

The magnetic part of $W$ takes the following form:
\begin{eqnarray}\label{WB}
B_{ab}:={^*} W_{a3b3}  & = & - {^*} W_{a33b}  \\
& = & -\frac12 \varepsilon_a{^{cd}}R_{3bcd}
 +\varepsilon_a{^{cd}} g_{c[d}S_{b]3} \\
& = & \varepsilon_a{^{cd}} K_{bc|d}
+\frac12 \varepsilon_a{^{cd}} h_{bc}S_{d3}\\
& = & \frac12 \varepsilon_a{^{cd}} K_{bc|d}
 + \frac12 \varepsilon_b{^{cd}} K_{ac|d}
\, ,
\end{eqnarray}
where $S_{3a} = R_{3a}$ is given by (\ref{R3a}),
and the magnetic part (\ref{WB}) we derived as follows:
\begin{eqnarray} \nonumber
 2 {^*} W_{a33b} &=&  \varepsilon_{{a3\mu\nu}}W^{\mu\nu}{_{3b}} =
  \varepsilon_{acd3}W^{cd}{_{3b}}=  \varepsilon_{acd}W^{cd}{_{3b}} =
 \varepsilon_a{^{cd}}W_{3bcd} \\
 &=& \varepsilon_a{^{cd}}R_{3bcd}
 +\varepsilon_a{^{cd}}g_{b[c}S_{d]3} - g{_{3[c}}S_{d]b}
 \, .\end{eqnarray}

The Einstein equations with cosmological constant
$$ \Lambda=-\frac3{l^2} \, $$
may be rewritten with the help of the conformal transformation \eq{CT} in the
following form:
\be\label{R33L} 0 = {\tilde R}_{33} - \Lambda{\tilde g}_{33} = R_{33}
 + \frac1z g_{ab} K^c{_c} \, , \ee
\be\label{RabL} 0 = {\tilde R}_{ab} - \Lambda{\tilde g}_{ab} = R_{ab}
+\frac2z K_{ab} +
 \frac1z g_{ab} K^c{_c} \, , \ee
 \be\label{R3aL} 0 = {\tilde R}_{3a} - \Lambda{\tilde g}_{3a} = R_{3a}
 = K^b{_{a|b}} - K^b{_{b|a}} \, , \ee
 where the last equality follows from (\ref{R3a}).
Equation (\ref{R3aL}) is the usual vector constraint. The scalar one
is hidden in the term $R_{ab}h^{ab}-R_{33}$. More precisely,
from (\ref{R33}), (\ref{Rab}), (\ref{R33L}) and (\ref{RabL}) we obtain
\[ \left[{\cal R}_{ab}(h) + K_{bc}K^c{_a}
    - K^c{_c} K_{ab}\right]h^{ab} +\frac4z K^c{_c}=0 \, .
\]

The equations (\ref{RabL}) imply the standard asymptotics (\ref{asc})
for the three-metric ${\gthree}$.
  Moreover,
for the extrinsic curvature $K_{ab}$ we get:
\be\label{asK}
K_{ab}= z\left({\stackrel{(2)}{h}} + 3z \ped \right) + O(z^3) \, ,\ee
where
\be\label{2h} {\stackrel{(2)}{h}}_{ab} =
\frac 14 {\stackrel{(0)}{h}}_{ab}{\cal R}\left({\stackrel{(0)}{h}}\right)
 - {\cal R}_{ab}\left({\stackrel{(0)}{h}}\right) \ee
(cf. \cite{Skenderis}). In addition, equation (\ref{R33})
together with (\ref{R33L}) rewritten in the form
 \[ \partial_3 K^c{_c} -\frac1z K^c{_c} +K^a{_b}K^b{_a}=0\]
 imply
 \be\label{trchi} \ped_{ab}{\stackrel{(0)}{h}}{^{ab}}=0 \, .\ee
Let us also notice that the leading order term in the vector constraint
\[ K^b{_{a|b}} - K^b{_{b|a}} =0 \]
corresponding to ${\stackrel{(2)}{h}}$ in (\ref{asK})
is equivalent to the contracted Bianchi identity
\[ {\cal R}^b{_{a|b}} - \frac12{\cal R}^b{_{b|a}} =0 \, . \]
However, the next order term gives for $\ped$ the following constraint:
\be\label{ttchi} {\nabla_b(\stackrel{(0)}{h})} \ped{^b}{_a} = 0 \, . \ee
Equations (\ref{trchi}) and (\ref{ttchi}) express the fact that the tensor
$\ped$ which is not determined by asymptotic analysis is transverse traceless
with respect to the metric $\stackrel{(0)}{h}$.

Finally,
\be\label{Echi} E_{ab}= -3z \ped_{ab} + O(z^2) \, ,\ee
\begin{eqnarray} B_{ab}& = & \frac{z}2\left(
 \varepsilon_a{^{cd}} {\stackrel{(2)}{h}}_{bc|d}
 +  \varepsilon_b{^{cd}} {\stackrel{(2)}{h}}_{ac|d}\right) + O(z^2)\\
 &=& \label{Bb} -\frac{z}2\left(
 \varepsilon_a{^{cd}} {\cal R}_{bc|d}
 +  \varepsilon_b{^{cd}} {\cal R}_{ac|d}\right) + O(z^2) \, .
 \end{eqnarray}
Let us denote
\be\label{bR} \beta_{ab}:= \varepsilon_a{^{cd}} {\cal R}_{bc|d}
 +  \varepsilon_b{^{cd}} {\cal R}_{ac|d} \ee
 the leading order term in $B_{ab}$ which plays a similar role to
 the tensor $\ped$ in $E_{ab}$.
 Symmetric tensor $\beta$ is equivalent to the Cotton tensor\footnote{
 Three-dimensional counterpart of the Weyl tensor.}
 \be\label{ct}
 C_{abc}:= {\cal R}_{ab|c}- {\cal R}_{ac|b} +\frac14
 \left( h_{ac}{\cal R}_{|b}-h_{ab}{\cal R}_{|c}\right) \,
 \ee
 via the following relation between them:
\[ \beta_{ab}=\varepsilon_a{^{cd}}C_{bcd}=2\varepsilon_a{^{cd}}{\cal R}_{bc|d}
-\frac12\varepsilon_{ab}{^{c}}{\cal R}_{|c} \]
 which implies that for the usual asymptotic AdS spacetime,
 i.e. when metric $\stackrel{(0)}{h}$
 is conformally flat and its Cotton tensor vanishes,
 the tensor $\beta$ has to vanish as well.

 In general case, for non-trivial $\stackrel{(0)}{h}$,
 contracted Bianchi identities for Ricci tensor ${\cal R}_{ab}$ imply
 that the tensor $\beta$ has the same properties as $\ped$, i.e. is
 transverse traceless.

A generalization of some results presented in this Section to higher
dimension of the spacetime can be found in \cite{AASD}, \cite{HIM}.

\section{Symplectic structure on tube}\label{sf}
In \cite{GRGKij} the following theorem was proved:
\begin{Theorem}
 Field dynamics in a four-dimensional region $O$ is equivalent to
\begin{equation}
\delta \int_{\cal O} L  = - \frac 1{16 \pi}  \int_{\partial{\cal O}}
g_{kl} \;  \delta \Pi^{kl}
\ , \label{dL-overO}
\end{equation}
where $g_{kl}$ is the three-dimensional metric induced on the boundary
${\partial{\cal O}}$ by $g_{\mu\nu}$ and $\Pi$
is the extrinsic curvature (in A.D.M. densitized form) of ${\partial{\cal O}}$.
\end{Theorem}
This theorem shows the universality of the symplectic structure:
\[ \int_{\partial{\cal O}}   \delta \Pi^{kl} \wedge \delta g_{kl} \, . \]
In our case a boundary data on $S$ consists of the three-metric
${\tilde h}_{ab}$ and canonical A.D.M. momentum $\tilde Q^{ab}$ which
is related with extrinsic curvature ${\tilde K}_{ab}$ in the usual way:
\[ \tilde Q_{ab}= \sqrt{-\tilde h}\left( {\tilde K}_{ab} -
{\tilde h}^{cd}{\tilde K}_{cd}{\tilde h}_{ab} \right) \, .\]
Conformal rescaling of the three-metric and extrinsic curvature
\[ {\tilde h}_{ab}=\Omega^{-2}2 h_{ab} \, , \quad
   {\tilde K}_{ab}= -\frac{\tilde\Gamma^3_{ab}}{\sqrt{{\tilde h}^{33}}}=
    \Omega^{-1} \left( { K}_{ab}-\frac1z h_{ab}\right) \]
enables one to analyze the symplectic structure as follows:
\begin{eqnarray} \tilde h_{ab}\delta {\tilde Q}^{ab} &= & \delta \left(
\tilde h_{ab} {\tilde Q}^{ab} \right) -
{\tilde Q}^{ab}\delta \tilde h_{ab} \nonumber \\
&=&
\delta \left(
\tilde h^{ab} {\tilde Q}_{ab} \right) +
{\tilde Q}_{ab}\delta \tilde h^{ab} \, . \label{hQ1}
\end{eqnarray}
In particular, (\ref{hQ1}) implies
\[ \int_{S}
 \delta {\tilde h}_{ab}\wedge \delta \tilde Q^{ab} =
\int_{S}
\delta \tilde Q_{ab} \wedge \delta {\tilde h}^{ab} \, .\]
Moreover,
\begin{eqnarray} \label{Qdg} \tilde Q_{ab}\delta {\tilde h}^{ab} &=&
  \Omega^{-2} \sqrt{-h}\left( {K}_{ab}-\frac2z h_{ab}-K^c{_c}h_{ab}\right)\delta {h}^{ab}\\
  & = & 4 \Omega^{-3} \frac1l\delta \sqrt{-h} +\Omega^{-2} \sqrt{-h}
  \left( {K}_{ab} - {h}^{cd}{K}_{cd}{h}_{ab} \right) \delta {h}^{ab} \, .
\end{eqnarray}
With the help of standard variational identitites:
\[ \delta \sqrt{-h} = \frac12 \sqrt{-h} h^{ab}\delta h_{ab} \, ,\]
\[ \delta R_{ab}(h) = \delta\Gamma^c{_{ab|c}} - \delta\Gamma^c{_{ac|b}} \, ,\]
\[ \delta \left( \sqrt{-h} R(h) \right)=
   \sqrt{-h}\left( R_{ab}-\frac12 h_{ab} R \right) \delta h^{ab} +\partial_c
   \left[ \sqrt{-h} \left( h^{ab}\delta\Gamma^c{_{ab}}
   - h^{ac}\delta\Gamma^b{_{ab}} \right) \right] \, ,\]
we analyze the singular part of (\ref{Qdg}) as follows:
\begin{eqnarray} \nonumber
{\rm sing}(\tilde Q_{ab}\delta {\tilde h}^{ab}) &=&
\delta\left( 4 \Omega^{-3} \frac1l \sqrt{-h}\right) + \sqrt{-h}\, \Omega^{-2}
  z \left( {\cal R}_{ab} - \frac12{h}_{ab}{h}^{cd}{\cal R}_{cd}
  \right) \delta {h}^{ab}  \\ \label{Qdgsing}
  &=& \frac1l \delta\left\{ 4 \Omega^{-3} \sqrt{-h} + \Omega^{-2} \sqrt{-h}
  {\cal R} \right\} + \mbox{full divergence}
\end{eqnarray}
which is a full variation up to boundary terms. Finally
\begin{eqnarray}
\lim_{\epsilon\rightarrow 0} \int_{S_\epsilon}
\delta \tilde Q_{ab} \wedge \delta {\tilde h}^{ab} &=& \nonumber
3l^2 \int_{{\scri}} \delta \left[\sqrt{-\det{\stackrel{(0)}{h}}}\left(\chi_{ab} -
{\stackrel{(0)}{h}}_{ab}\chi^c{_c} \right)\right]
\wedge \delta {\stackrel{(0)}{h}}{^{ab}} \\
&=&
3l^2 \int_{{\scri}} \delta \left[\sqrt{-\det{\stackrel{(0)}{h}}}
\left(\chi^c{_c} {\stackrel{(0)}{h}}{^{ab}} - \chi^{ab} \right)\right]
\wedge \delta {\stackrel{(0)}{h}}{_{ab}} \label{dQdg}
\, ,
\end{eqnarray}
where $S_\epsilon:= \{ z=x^3=\epsilon \}$ is a tube close to infinity.
Symplectic structure (\ref{dQdg}) on conformal boundary consists of
the metric ${\stackrel{(0)}{h}}{_{ab}}$ and canonically conjugated
momenta $\pi^{ab}:=3l^2\sqrt{-\det{\stackrel{(0)}{h}}}
\left(\chi^c{_c} {\stackrel{(0)}{h}}{^{ab}} - \chi^{ab} \right)$
in A.D.M. densitized form.

\section{Asymptotic charges}\label{asch}
Let
${\tilde F}_{\mu\nu}:= {\tilde W}_{\mu\nu\rho\sigma}{\tilde Q}^{\rho\sigma}$
and
${F}_{\mu\nu}:= {W}_{\mu\nu\rho\sigma}{Q}^{\rho\sigma}$ respectively.
The conformal rescaling (\ref{conf4}) and Theorem \ref{conf_resc_th}
(see \cite{JJML}) imply a simple
relation between ${\tilde F}$ and $F$:
\be\label{FCT}
{\tilde F}_{\mu\nu} = \Omega^{-1} {F}_{\mu\nu} \, .
\ee
According to \cite{cykem}, Hodge dual of the two-form $\tilde F$
represented by a bivector density
defines an integral quantity at $\scri$ as follows:
\[ I(C):= \lim_{z\rightarrow 0^+}
\int_{C_z}  {\tilde F}^{\mu\nu} \rd {\tilde S}_{\mu\nu} =
 \int_{C} \left(\lim_{z\rightarrow 0^+} \Omega^{-1} {F}^{\mu\nu}
  \rd S_{\mu\nu} \right) \, ,\]
where $C_z$ is a family of spheres approaching
$C$ -- sphere at infinity (cut of $\scri$),
\[ \rd {\tilde S}_{\mu\nu} :=
\sqrt{-\det\tilde g} \partial_\mu \wedge\partial_\nu \rfloor
 \rd z^0\wedge\ldots\wedge\rd z^3 \, ,\]
\[ \rd {S}_{\mu\nu} :=
\sqrt{-\det g} \partial_\mu \wedge\partial_\nu \rfloor
 \rd z^0\wedge\ldots\wedge\rd z^3 \, .\]

Let us consider asymptotic CYK tensor 
as a two-form $\tilde Q$ such that a boundary value at $\scri$ of
the corresponding rescaled tensor $Q$ is a linear combination of
   (\ref{asQ}) and (\ref{asQ*}),
 i.e. its boundary value is the same as in the case of
 pure AdS spacetime (\ref{gAdS}).

 For asymptotic AdS spacetime formulae (\ref{Echi}) and (\ref{Bb}) imply that
 $\lim_{z\rightarrow 0^+} \Omega^{-1} W$ is finite.
 Moreover, for a given ACYK tensor $\tilde Q$ we obtain the well
 defined expression $I(C)$ which depends only on asymptotic values at $\scri$.
Let us check that for a given value $Q$ at $\scri$ the quantity $I(C)$
does not depend on the choice of cut $C$, i.e. represents a conserved quantity.
We have
\[ \int_{C} \left(\lim_{z\rightarrow 0^+} \Omega^{-1} {F}^{\mu\nu}
  \rd S_{\mu\nu} \right)= \int_{C} \left(\lim_{z\rightarrow 0^+}
  \Omega^{-1} {F}^{3a} \rd S_{a} \right) \, ,\]
where $\rd S_{a}:=\partial_a \rfloor {\rm vol}({\stackrel{(0)}{h}})$.
Moreover, for $Q(L)= {\stackrel{(0)}{h}}(L)\wedge\rd z$ (cf. (\ref{asQ}))
\[ \lim_{z\rightarrow 0^+} \Omega^{-1} {F}^{3a}=\lim_{z\rightarrow 0^+}
  \Omega^{-1} {W}^{3a\mu\nu}Q_{\mu\nu}= 6l \chi^a{_b}L^b \, .\]
 Similarly, for $\ast{}Q(L) =  L \rfloor {\rm vol}({\stackrel{(0)}{h}})$
 (cf. (\ref{asQ*})) we get
\[ \lim_{z\rightarrow 0^+} \Omega^{-1} {F}^{3a}=\lim_{z\rightarrow 0^+}
  \Omega^{-1} {W}^{3a\mu\nu}\ast\! Q_{\mu\nu}= l \beta^a{_b}L^b \, ,\]
where $\beta^a{_b}$ is given by (\ref{bR}).
If $L=L^b\partial_b$ is a conformal Killing vector field for the metric
${\stackrel{(0)}{h}}$ (which is true for ACYK tensors $Q$)
 the conservation law for $I(C)$
results from transverse traceless property of tensors $\chi$ and $\beta$.
More precisely, for three-volume $V\subset\scri$ such that
$\partial V=C_1\cup C_2$ we have
\begin{eqnarray}
 \int_{C_1} \chi^a{_b}L^b \rd S_{a} - \int_{C_2} \chi^a{_b}L^b \rd S_{a}&=&
\int_{\partial V} \chi^a{_b}L^b \rd S_{a} =
\int_{V} \nabla_a(\chi^a{_b}L^b){\rm vol}({\stackrel{(0)}{h}}) \\
& = &
\int_{V} \left[ L^b\nabla_a\chi^a{_b} +\chi^{ab}L_{(a|b)}
\right]{\rm vol}({\stackrel{(0)}{h}}) =0 \, .
\end{eqnarray}
Let us define the following quantity:
\be\label{HQ} H(Q):=\frac{l}{32\pi} \int_{C}  \Omega^{-1} {F}^{\mu\nu}(Q)
  \rd S_{\mu\nu} \, .\ee
For ACYK tensor $\tilde Q$ in asymptotic AdS spacetime
the corresponding quantity $H(Q)$ is conserved,
i.e. does not depend on the choice of spherical cut $C$.
In particular, for the conformal Killing vector field $L$ and $Q(L)$
given by (\ref{asQ}) the conserved charge $H(Q(L))$ may be expressed
in terms of electric part of Weyl tensor and takes
the following form\footnote{The apparent incompatibility of factors
between our integral $H(Q(L))$ and Ashtekar's definition is related
to a different choice of conformal factor $\Omega$, our choice is twice
smaller.}
proposed by Ashtekar \cite{AM}, \cite{AASD} (see also \cite{HIM}):
\be\label{EL}
 H(Q(L))= -\frac{l}{16\pi} \int_{C} \Omega^{-1} E^a{_b}L^b \rd S_{a} \, .\ee
In the Schwarzschild-AdS spacetime (\ref{SAdS}) for the Killing vector
\be\label{ddt} L ={\partial\over\partial t}= l^{-1}
{\partial\over\partial \bar t}=l^{-1}\partial_0 \ee
 definition (\ref{HQ}) gives (minus) mass:
\be\label{mass} H(Q(L))=
-\frac{1}{16\pi} \int_{C} \Omega^{-1} E^a{_0} \rd S_{a} =
  \frac{3l}{16\pi} \int_{C} \chi^0{_0}
  \sqrt{-\det {\stackrel{(0)}{h}}}\rd\theta\rd\phi =  - m \, .\ee
The last equality in the above formula follows from
(\ref{h0S}) and (\ref{chiS}).
Obviously, the same value $-m$ we obtain
 for Kerr-AdS metric (\ref{KerrAdSg}). Moreover, in the Kerr-AdS spacetime
for $L= {\partial\over\partial \phi}$ we obtain the angular momentum:
\be\label{angmom}
 H(Q(L))= -\frac{l}{16\pi} \int_{C} \Omega^{-1} E^a{_\phi} \rd S_{a} =
  \frac{3l^2}{16\pi} \int_{C} \chi^0{_\phi}
  \sqrt{-\det {\stackrel{(0)}{h}}}\rd\theta\rd\phi =   ma \, . \ee
The details of calculations for the Kerr-AdS spacetime we present
in the Appendix \ref{mam}. Let us observe that our conserved quantity
$H(Q(L))$ in terms of the
symplectic momenta $\pi^{ab}$ at $\scri$ takes the following form:
\be\label{piL}
 H(Q(L))= -
  \frac{1}{16\pi} \int_{C} \pi^0{_b}L^b \rd\theta\rd\phi  \, , \ee
which is in the same A.D.M. form as the usual linear
or angular momentum
at spatial infinity in asymptotically flat spacetime
(cf. \cite{CJK} p. 80).\\
\underline{Remark:} In general case, when ${\stackrel{(0)}{h}}$
 is not conformally flat,
 it may happen that one obtains asymptotic charge which is no longer
 conserved --- Bondi-like phenomena (cf. \cite{cykem}).

The ``topological'' charge one can try to define as follows:
\[ H(*Q(L))= \frac{l}{32\pi} \int_{S^2}  \Omega^{-1} {F}^{\mu\nu}(*Q)
  \rd S_{\mu\nu} =
 -\frac{l}{16\pi} \int_{S^2} \Omega^{-1} B^a{_b}L^b \rd S_{a} \, .\]
We want to stress that, in general, we can meet problems with
finding spherical cuts of $\scri$. Hence the choice of a domain of integration
for the corresponding two-form $\Omega^{-1} {F}^{\mu\nu}(*Q)
  \rd S_{\mu\nu}$ has to be carefully analyzed.
 In NUT-AdS spacetime a conformal boundary $\scri$
equipped with the metric (\ref{h0NUT}) is a
non-trivial bundle over $S^2$ -- two-dimensional sphere.
However, for $L$ given by (\ref{ddt}),
when the above formula
pretends to define ``dual mass'' charge, we have
\begin{eqnarray} - \Omega^{-1} B^a{_b}L^b \rd S_{a} &=&
  -\frac{1}{l} \Omega^{-1} B^a{_0} \rd S_{a} \\ &=&
   \frac{1}{2} \sqrt{-\det \stackrel{(0)}{h}}
   \left[ \beta^0{_0}\rd\theta\wedge \rd\phi
   + \rd {\bar t} \wedge ( \beta^\phi{_0}\rd\theta
   - \beta^\theta{_0} \rd\phi   ) \right] \nonumber \\ &=&
  2{\bar l}\sin\theta\rd\theta\wedge\rd\phi \, ,
   \end{eqnarray}
where the last equality one can easily check using formulae from Appendix B.
Let us notice that the resulting two-form projects uniquely
on the base manifold which is a two-dimensional sphere.
 Finally we have
\[ 2H(*Q(L))= \frac{l}{16\pi} \int_{S^2} 4 {\bar l}\sin\theta\rd\theta\rd\phi
   =  l {\bar l} =  {\bf l} \]
which confirms that we can interpret the NUT parameter $\bf l$ as
a dual mass charge.
\section{Conclusions}
We have constructed all solutions to CYK equation in AdS (and de-Sitter)
spacetime via pullback technique from five-dimensional flat ambient space.

Analyzing three important examples: Schwarzschild-AdS, Kerr-AdS and NUT-AdS,
we have shown how geometrically natural two-form $F$ (built from
Weyl tensor $W$ and ACYK tensor $Q$), leading to the universal definition
of a global charge (\ref{HQ}), enables one to understand energy,
angular momentum and dual mass in asymptotic AdS spacetime.
Definition (\ref{HQ}) occurred to be equivalent to two other important
formulae (\ref{EL}) and (\ref{piL}).

The relation between Killing vector fields $L$ and CYK tensors $Q$ has been
examined. However, the relation (\ref{l12}) between AdS and Minkowski suggests
some ambiguity in the definition of angular momentum when the three-metric
${\stackrel{(0)}{h}}$ is not conformally flat.
More precisely, CYK tensor ${}^{[12]} Q - *{}^{[34]} Q$
corresponds to CYK tensor $\widetilde{\cal L}_{12}$ in Minkowski
hence $H({}^{[12]} Q - *{}^{[34]} Q)$ should correspond to the
third component of the angular momentum.
Obviously, in the standard asymptotic AdS spacetime the quantity
$H(*{}^{[34]} Q)$ vanishes and the ambiguity disappears.
\appendix
\section{Canonical coordinates for AdS-Kerr near $\scri$}
The solution of Einstein equations with mass, angular momentum
and negative cosmological constant explicitly given by
(\ref{KerrAdSg}) can be rewritten near $\scri$ as follows:
\begin{eqnarray}
{\tilde g}_{\mbox{\tiny\rm Kerr-AdS}} & = & \nonumber
 \frac{l^2}{w^2}(1+{\bar a}^2 w^2\cos^2\theta)\left(
 \frac{\rd w^2}{1+w^2(1+{\bar a}^2-bw+{\bar a}^2 w^2)}
+ \frac{\rd \theta^2}{1-{\bar a}^2\cos^2\theta} \right) \\ \label{KerrAdS2}
& & + {\tilde g}_{tt} \rd t^2 + 2{\tilde g}_{t\phi} \rd t \rd \phi +
 {\tilde g}_{\phi\phi} \rd \phi^2   ,
\end{eqnarray}
where ${\bar a}:=\frac{a}l$, $b:=\frac{2m}l$, $w:=\frac{l}r$ and
\[
{\tilde g}_{tt} = -\frac1{w^2}-1-{\bar a}^2\sin^2\theta
 +{bw \over 1+{\bar a}^2w^2\cos^2\theta} \, , \]
 \[
 {\tilde g}_{t\phi} = l{\bar a}\sin^2 \theta \left( \frac{1}{w^2}
+{\bar a}^2 - {bw \over 1+{\bar a}^2 w^2\cos^2\theta}\right) \, , \]
\begin{equation}\label{KerrAdScomp2}
{\tilde g}_{\phi\phi} = l^2\sin^2\theta
\left[\left(\frac1{w^2}+{\bar a}^2\right)(1-{\bar a}^2)+
{bw{\bar a}^2\sin^2\theta\over 1+{\bar a}^2 w^2\cos^2\theta}\right] .
\end{equation}
The canonical coordinate $z(w,\theta)$ fulfilling eikonal equation:
\be\label{ez} \left\|  \frac{l\rd z}z \right\|^2 = 1 \ee
can be found with the help of the following conditions:
\be\label{wz} \left(\frac{w}z \frac{\partial z}{\partial w}\right)^2
    [1+w^2(1+{\bar a}^2-bw+{\bar a}^2 w^2)]=1 \, ,\ee
\be\label{tz} \left(\frac1z\frac{\partial z}{\partial\theta}\right)^2
    (1-{\bar a}^2\cos^2\theta)= {\bar a}^2\cos^2\theta \, .\ee
It is easy to check that each solution of (\ref{wz}-\ref{tz})
is simultaneously a solution of (\ref{ez}) for the metric (\ref{KerrAdS2}).
We are looking for a solution of (\ref{wz}-\ref{tz}) in the following form:
\[ \ln z = A(w) + B(\theta) \]
which leads to the following ODE's:
\be\label{wzA} w \frac{\rd A}{\rd w}=\frac1{\sqrt{
    1+w^2(1+{\bar a}^2-bw+{\bar a}^2 w^2)}} \ee
\be\label{tzB} \frac{\rd B}{\rd \theta}=
     \frac{{\bar a}\cos\theta}{\sqrt{1-{\bar a}^2\cos^2\theta}} \, .\ee
The solution of equation (\ref{wzA}) expresses in terms of the elliptic integral
\[ A = \int \frac{\rd w}{w\sqrt{1+w^2(1+{\bar a}^2-bw+{\bar a}^2 w^2)}}  \]
but (\ref{tzB}) possesses the simple solution:
\[ B= \ln \left(
{\bar a}\sin\theta+{\sqrt{1-{\bar a}^2\cos^2\theta}}\right) \, . \]
Moreover, considerations (similar to (\ref{zw}) in the case of
 Schwarzschild-AdS metric)
lead to the following formula:
\[ A = \ln \left( \frac{w}{1+\sqrt{1+w^2(1+{\bar a}^2)}}\right) +
{\bar F}({\bar a}, b, w) \]
with the function $\bar F$ implicitly defined by the following
conditions:
\[ {\bar F}({\bar a}, b, 0)=0 \, , \quad {\bar F}(0,b,w)= b F(b,w) \, ,\]
\be\label{dFdw} w\frac{\partial {\bar F}}{\partial w} =
\frac1{\sqrt{1+w^2(1+{\bar a}^2-bw+{\bar a}^2 w^2)}} -
\frac1{\sqrt{1+w^2(1+{\bar a}^2)}} \, .\ee
Finally, we have
\be\label{zwt} z(w,\theta) = \frac{w}{1+\sqrt{1+w^2(1+{\bar a}^2)}}\left(
{\bar a}\sin\theta+{\sqrt{1-{\bar a}^2\cos^2\theta}}\right)
\exp[{\bar F}({\bar a},b,w)]\, . \ee
Analyzing series expansion of the right-hand side of (\ref{dFdw})
we can easily produce an asymptotic form of $\bar F$:
\be\label{asF} {\bar F}= \frac{b}6 w^3 -\frac18{\bar a}^2 w^4
-\frac{3b}{20}(1+{\bar a}^2)w^5 +
\frac1{16}[ 2{\bar a}^2 (1+{\bar a}^2) +b^2] w^6 + O(w^7) \ee
which enables one to analyze asymptotics at $\scri$.
However, to obtain the canonical form (\ref{canform})
for the metric ${\tilde g}_{\mbox{\tiny\rm Kerr-AdS}}$ we should
change a coordinate
$\theta$ because $\tilde g(\!\rd z,\!\rd\theta)$ is not vanishing.
We are looking for a new coordinate ${\bar\theta}(w,\theta)$
with the following properties:
\[ \bar\theta(0,\theta)=\theta \, , \quad
\tilde g(\rd z, \rd\bar\theta)= 0 \, .\]
Orthogonality of coordinates $z$ and $\bar\theta$
leads to the following condition:
\[ \frac{\partial \bar\theta}{\partial w}
+\frac{{\bar a} w\cos\theta\sqrt{1-{\bar a}^2\cos^2\theta}}
{\sqrt{1+w^2(1+{\bar a}^2-bw+{\bar a}^2 w^2)}}
\frac{\partial \bar\theta}{\partial\theta} = 0 \]
which implies that the curve $\bar\theta(w,\theta)=$ const. obeys
the following ODE:
\be\label{dwdth}
\frac{\rd\theta}{\rd w}= - { \frac{\partial \bar\theta}{\partial w}
\over\frac{\partial \bar\theta}{\partial\theta} } =
 \frac{{\bar a} w\cos\theta\sqrt{1-{\bar a}^2\cos^2\theta}}
{\sqrt{1+w^2(1+{\bar a}^2-bw+{\bar a}^2 w^2)}} \, .
\ee
It is easy to verify that ODE (\ref{dwdth}) leads again to elliptic integral.
More precisely, let us define
\[ \delta(\tau) := {\bar a}
\int_0^\tau \frac{w\rd w}{\sqrt{1+w^2(1+{\bar a}^2-bw+{\bar a}^2 w^2)}}\, ,
 \]
then the solution of (\ref{dwdth}) takes the following form:
\begin{eqnarray}\label{datt}
\delta(w) & = & \artanh \frac{\sin\theta}{\sqrt{1-{\bar a}^2\cos^2\theta}}
   - \artanh \frac{\sin\bar\theta}{\sqrt{1-{\bar a}^2\cos^2\bar\theta}}
   \\ & = &
  {\bar a} \left[ \frac12 w^2 -\frac18 (1+{\bar a}^2)w^4 +\frac{b}{10}w^5
  + O(w^6) \right] \, .
  \end{eqnarray}
Now, we are ready to calculate induced three-metric $h$ on
the surface $S=\left\{ z = \mbox{const.} \right\}$.
We have
\[ 0= \rd A +\rd B= \frac{\rd w}{w\sqrt{1+w^2(1+{\bar a}^2-bw+{\bar a}^2 w^2)}}
 + \frac{{\bar a}\cos\theta}{\sqrt{1-{\bar a}^2\cos^2\theta}} {\rd \theta}\]
and (from (\ref{datt}))
\[ \frac{\rd\theta}{\cos\theta\sqrt{1-{\bar a}^2\cos^2\theta}}
   - \frac{\rd\bar\theta}{\cos\bar\theta\sqrt{1-{\bar a}^2\cos^2\bar\theta}} =
    \frac{{\bar a}w\rd w}{\sqrt{1+w^2(1+{\bar a}^2-bw+{\bar a}^2 w^2)}} \]
which implies
\be\label{thbarth}  \frac{\rd\theta}{\cos\theta\sqrt{1-{\bar a}^2\cos^2\theta}}
     (1+{\bar a}^2w^2\cos^2\theta )
   = \frac{\rd\bar\theta}{\cos\bar\theta\sqrt{1-{\bar a}^2\cos^2\bar\theta}}
  \, .  \ee
Let us define $\bar t:= \frac{t}l$.
Formulae (\ref{datt}) and (\ref{thbarth}) together with (\ref{KerrAdS2})
enable one to derive implicitly the induced three-metric $h$ with respect
to coordinates $\bar t, \bar\theta, \phi$:
\begin{eqnarray} \nonumber
\frac{z^2}{l^2}{\tilde g}_{\mbox{\tiny\rm Kerr-AdS}}\Big|_S & = &
\frac{z^2}{w^2}\Big\{ \frac{\cos^2\theta}{\cos^2\bar\theta
(1-{\bar a}^2\cos^2\bar\theta)}{\rd\bar\theta^2} + \\ & & + \nonumber
2{\bar a}\sin^2\theta\left( 1+{\bar a}^2 w^2 -
\frac{bw^3}{1+{\bar a}^2w^2\cos^2\theta} \right)\rd{\bar t}\rd\phi \\ & &
+\sin^2\theta \left[ (1+{\bar a}^2w^2)(1-{\bar a}^2) + \label{hmetric}
\frac{bw^3{\bar a}^2\sin^2\theta}{1+{\bar a}^2w^2\cos^2\theta}\right] \rd\phi^2
\\ & & - \left(1+{\bar a}^2w^2\sin^2\theta + w^2-
\frac{bw^3}{1+{\bar a}^2w^2\cos^2\theta}\right)\rd{\bar t}^2
\Big\} \nonumber \, .
\end{eqnarray}
Let us observe that the conformal factor
\be\label{zw1} \frac{z}w=\frac{
{\bar a}\sin\theta+{\sqrt{1-{\bar a}^2\cos^2\theta}}}
{1+\sqrt{1+w^2(1+{\bar a}^2)}} \exp{\bar F}
\ee
is regular at $\scri$ (corresponding to surface $\{ w=0 \}$).
To finish derivation of asymptotics of $\gthree$
we have to notice that (\ref{datt}) written in equivalent form
as follows:
\be\label{btwt}
 \frac{\sin\bar\theta}{\sqrt{1-{\bar a}^2\cos^2\bar\theta}}=
\frac{\sin\theta -\tanh\delta(w)\sqrt{1-{\bar a}^2\cos^2\theta}}
{\sqrt{1-{\bar a}^2\cos^2\theta}-\tanh\delta(w)\sin\theta}
\ee
defines (implicitly) the function $\bar\theta(\theta,w)$.
In particular
\[ \cos^2\theta - \cos^2\bar\theta =
   -2\delta\cos^2\bar\theta\sin\bar\theta\sqrt{1-{\bar a}^2\cos^2\bar\theta}
   + O(\delta^2) \]
which is a straightforward consequence of (\ref{btwt}).

Equations (\ref{zwt}) and (\ref{btwt}) define
the mapping $(w,\theta)\mapsto (z,{\bar\theta})$.
To obtain the explicit form of $\gthree$
this mapping should be inverted in the neighbourhood of $z=w=0$
corresponding to $\scri$ and applied to (\ref{hmetric}).

If we put $w=0$ in (\ref{hmetric}), we obtain
the asymptotic value of $\gthree$ at $\scri$:
\begin{eqnarray} {\stackrel{(0)}{h}}& = & \frac14\left(
{\bar a}\sin\bar\theta+{\sqrt{1-{\bar a}^2\cos^2\bar\theta}}\right)^2
\Big[ \frac{1}{1-{\bar a}^2\cos^2\bar\theta}{\rd\bar\theta^2} + \label{h0Kerr}
\\ & & + \nonumber
2{\bar a}\sin^2\bar\theta \rd{\bar t}\rd\phi 
+\sin^2\bar\theta (1-{\bar a}^2) \rd\phi^2 - \rd{\bar t}^2 \Big] \, .
\end{eqnarray}
\noindent \underline{Remark}:
Let us notice that sometimes the coordinates $({\bar t}, w,\theta,\phi)$ are not
convenient in the asymptotic region. According to \cite{HT} one can
introduce new spatial coordinates $(\rho,\Theta,\Phi)$ defined as follows:
\begin{eqnarray} \label{APhi}
\Phi &:= & (1-{\bar a}^2)\phi + {\bar a}{\bar t} \, ,\\
\rho^{-1}\cos\Theta & = & w^{-1}\cos\theta \, ,\\
(1-{\bar a}^2)\rho^{-2} &=& w^{-2} +{\bar a}^2 \sin^2\theta
- {\bar a}^2w^{-2}\cos^2\theta \, . \label{AR}
\end{eqnarray}
Some useful formulae describing coordinate transformation
$(w,\theta,\phi) \leftrightarrow (\rho,\Theta,\Phi)$
are given in Appendix \ref{HTc}. In particular,
they enable one to prove that
in new coordinates the induced metric $\gthree$ at $\scri$ takes
the following form:
\begin{eqnarray} {\stackrel{(0)}{h}}& = &
\frac{1-{\bar a}^2}{4\left(1-{\bar a}\sin\Theta\right)^2}
\Big[ {\rd\Theta^2} +\sin^2\Theta \rd\Phi^2 - \rd{\bar t}^2 \Big] \, ,
\end{eqnarray}
which explicitly shows that metric ${\stackrel{(0)}{h}}$ (for $\bar a < 1$)
 is in the conformal class of the
Einstein static universe (cf. (\ref{Esu})).

Let us denote by $\omega$ the following function:
 \[ \omega (\bar\theta) := \frac2{
{\bar a}\sin\bar\theta+{\sqrt{1-{\bar a}^2\cos^2\bar\theta}}}
\]
which comes from conformal factor in metric tensor ${\stackrel{(0)}{h}}$
given by (\ref{h0Kerr}).
To derive higher order terms in ${\gthree}_{ab}$ for Kerr-AdS we have to
check the following formulae:
\begin{eqnarray}
\frac{\omega(\bar\theta)}{\omega(\theta)} =
\frac{{\bar a}\sin\theta+{\sqrt{1-{\bar a}^2\cos^2\theta}}}
{{\bar a}\sin\bar\theta+{\sqrt{1-{\bar a}^2\cos^2\bar\theta}}} \nonumber
& = & 1 + {\bar a}\cos^2{\bar\theta} \, \delta + O(\delta^2) \, ,
\end{eqnarray}
\begin{eqnarray} \frac{\cos^2\theta}{\cos^2\bar\theta}& = & 1-2\delta
\sin\bar\theta{\sqrt{1-{\bar a}^2\cos^2\bar\theta}} +O(\delta^2) \nonumber
\\ & = &  \label{cos2}
1-{\bar a}\sin\bar\theta{\sqrt{1-{\bar a}^2\cos^2\bar\theta}}\, w^2 +O(w^4)\, , \\
\frac{\sin^2\theta}{\sin^2\bar\theta}& = & \label{sin2}
1+{\bar a}\frac{\cos^2\bar\theta}{\sin\bar\theta}
{\sqrt{1-{\bar a}^2\cos^2\bar\theta}}\, w^2 +O(w^4) \, ,
\end{eqnarray}
which together with (\ref{zw1}) and (\ref{asF}) imply
\be\label{zw2}
\left(\frac{z}w\right)^2 = \omega(\bar\theta)^{-2}
\left[1- (1+{\bar a}^2\sin^2\bar\theta)w^2+\frac{b}3 w^3 +O(w^4) \right] \, .
\ee
Now we are ready to derive higher order terms in (\ref{hmetric}).
We obtain the following non-vanishing components of three-metric $h$:
\begin{eqnarray}
h_{\bar\theta\bar\theta} & = & \nonumber
\frac{z^2}{w^2} \frac{\cos^2\theta}{\cos^2\bar\theta}
\frac1{(1-{\bar a}^2\cos^2\bar\theta)} \\
 &= &\frac{ 1- (1+{\bar a}^2\sin^2\bar\theta)w^2
 - {\bar a}\sin\bar\theta{\sqrt{1-{\bar a}^2\cos^2\bar\theta}}\, w^2
+\frac{b}3 w^3 +O(w^4)}{(1-{\bar a}^2\cos^2\bar\theta)
\omega(\bar\theta)^{2}} \, ,\\
h_{\bar t \bar t} & = & \nonumber -\frac{z^2}{w^2}
\left(1+{\bar a}^2w^2\sin^2\theta + w^2-
\frac{bw^3}{1+{\bar a}^2w^2\cos^2\theta}\right) \\
& = & - \omega(\bar\theta)^{-2}
\left[1- \frac{2b}3 w^3 +O(w^4) \right] \, ,\\
h_{\bar t\phi} & = &
{\bar a}\sin^2\theta \frac{z^2}{w^2}\left( 1+{\bar a}^2 w^2 -
\frac{bw^3}{1+{\bar a}^2w^2\cos^2\theta} \right)=
{\bar a}\sin^2\bar\theta \omega(\bar\theta)^{-2} \, \times  \\
& \times & \nonumber
\left[1+{\bar a}\frac{\cos^2\bar\theta}{\sin\bar\theta}
{\sqrt{1-{\bar a}^2\cos^2\bar\theta}}\, w^2
 - (1-{\bar a}^2\cos^2\bar\theta)w^2-\frac{2b}3 w^3 +O(w^4) \right]\, ,\\
h_{\phi\phi}&= & \nonumber
\sin^2\theta \frac{z^2}{w^2}\left[ (1+{\bar a}^2w^2)(1-{\bar a}^2) +
\frac{bw^3{\bar a}^2\sin^2\theta}{1+{\bar a}^2w^2\cos^2\theta}\right]
\\ &=& \sin^2\bar\theta \omega(\bar\theta)^{-2} (1-{\bar a}^2)
\left[ 1+
 \frac{{\bar a}^2\sin^2\bar\theta}{1-{\bar a}^2}bw^3 +O(w^4) \right]
 \times \\ &\times & \nonumber
\left[1+{\bar a}\frac{\cos^2\bar\theta}{\sin\bar\theta}
{\sqrt{1-{\bar a}^2\cos^2\bar\theta}}\, w^2
 - (1-{\bar a}^2\cos^2\bar\theta)w^2+\frac{b}3 w^3 +O(w^4) \right] \, .
\end{eqnarray}
Let us observe that asymptotics at $\scri$ with respect to coordinate
$w$ with coefficients depending on $\bar\theta$ can be easily transformed
into demanded asymptotics with respect to coordinate $z$.
More precisely,  from (\ref{zw2}) we have $z\omega(\bar\theta)=w(1+O(w^2))$
which implies
\[ a_0(\bar\theta) + a_2(\bar\theta) w^2 + a_3(\bar\theta) w^3 + O(w^4)=
    a_0(\bar\theta) + a_2(\bar\theta)\omega(\bar\theta)^2 z^2
    + a_3(\bar\theta) \omega(\bar\theta)^3 z^3 + O(z^4) \, ,\]
which means that the coefficients in the third degree polynomial
do not mix each other when we pass from $w$ to $z$.\\
Hence, we get
\begin{eqnarray} h &=& \nonumber
\frac{ \omega^{-2}
- (1+{\bar a}^2\sin^2\bar\theta) z^2
 - {\bar a}\sin\bar\theta{\sqrt{1-{\bar a}^2\cos^2\bar\theta}} z^2
+\frac{b}3 \omega z^3}{(1-{\bar a}^2\cos^2\bar\theta)}
 \rd\bar\theta^2  \\
& & +\, 2 {\bar a}\sin^2\bar\theta
\left[\frac1{\omega^{2}}+{\bar a}\frac{\cos^2\bar\theta}{\sin\bar\theta}
{\sqrt{1-{\bar a}^2\cos^2\bar\theta}}\, z^2
 - (1-{\bar a}^2\cos^2\bar\theta)z^2-\frac{2b}3\omega z^3 \right]
 \!\rd\bar t \rd\phi \nonumber \\ \nonumber
 & & + \sin^2\bar\theta  (1-{\bar a}^2)
\left[ 1+
 \frac{{\bar a}^2\sin^2\bar\theta}{1-{\bar a}^2}b \omega^3 z^3 \right]
 \times \\ \nonumber & & \times
\left[\omega^{-2} +{\bar a}\frac{\cos^2\bar\theta}{\sin\bar\theta}
{\sqrt{1-{\bar a}^2\cos^2\bar\theta}}\, z^2
 - (1-{\bar a}^2\cos^2\bar\theta)z^2+\frac{b}3 \omega z^3 \right] \rd\phi^2 \\
 & & -\left[\omega^{-2}
- \frac{2b}3 \omega z^3  \right] \rd{\bar t}^2 +O(z^4) \, ,
\end{eqnarray}
\begin{eqnarray} {\stackrel{(2)}{h}} & = & \nonumber
-\frac{(1+{\bar a}^2\sin^2\bar\theta)
 + {\bar a}\sin\bar\theta{\sqrt{1-{\bar a}^2\cos^2\bar\theta}}}
 {(1-{\bar a}^2\cos^2\bar\theta)}
 \rd\bar\theta^2  \\
& & + \,  2 {\bar a}\sin^2\bar\theta
\left[{\bar a}\frac{\cos^2\bar\theta}{\sin\bar\theta}
{\sqrt{1-{\bar a}^2\cos^2\bar\theta}}
 - (1-{\bar a}^2\cos^2\bar\theta) \right]
 \rd\bar t \rd\phi \nonumber \\ 
 & & + \, \sin^2\bar\theta  (1-{\bar a}^2)
\left[{\bar a}\frac{\cos^2\bar\theta}{\sin\bar\theta}
{\sqrt{1-{\bar a}^2\cos^2\bar\theta}}
 - (1-{\bar a}^2\cos^2\bar\theta) \right] \rd\phi^2 \label{h2} \, ,
\end{eqnarray}
\begin{eqnarray} \label{chiKerr}
  \ped & = & \nonumber  \frac{b\omega}3 \Big\{ 2\rd{\bar t}^2
  -4 {\bar a}\sin^2\bar\theta
 \rd\bar t \rd\phi + \left(1-{\bar a}^2+{3{\bar a}^2\sin^2\bar\theta}
   \right) \sin^2\bar\theta \rd\phi^2 \\
 & & + \frac{\rd\bar\theta^2}{1-{\bar a}^2\cos^2\bar\theta} \Big\} \, .
 \end{eqnarray}
\subsection{Mass and angular momentum of Kerr-AdS}\label{mam}
From (\ref{h0Kerr}) we have
\[ \sqrt{-\det \stackrel{(0)}{h}} = \omega^{-3}\sin\theta  \, .\]
Moreover, (\ref{chiKerr}) implies
\[ \ped^0{_\phi} =  b {\bar a} \omega^3 \sin^2\theta = \frac{2ma}{l^2}
 \omega^3 \sin^2\theta \, .\]
Hence for cut $C:=\{ t=\mbox{const.}\}\subset\scri$ we obtain
\[ \frac{3l^2}{16\pi}\int_{C} \ped^0{_\phi} \sqrt{-\det \stackrel{(0)}{h}}
  \rd\theta\rd\phi= \frac{3ma}{4}\int_0^\pi \sin^3\theta\rd\theta = ma \, \]
which gives (\ref{angmom}).
To obtain (\ref{mass}) we observe that
\[ \ped^0{_0} = -\frac23 b \omega^3 = -\frac{4m}{3l} \omega^3 \]
which implies
\[ \frac{3l}{16\pi}\int_{C} \ped^0{_0} \sqrt{-\det \stackrel{(0)}{h}}
  \rd\theta\rd\phi=
  -\frac{m}{2}\int_0^\pi \sin\theta\rd\theta = - m \, .\]

\subsection{``Henneaux-Teitelboim'' coordinates $\rho,\Theta,\Phi$
in Kerr-AdS}\label{HTc}
The following formulae:
\begin{eqnarray}
\rho^2 & = & {(1-{\bar a}^2)w^2\over
 1- {\bar a}^2\cos^2\theta +{\bar a}^2w^2\sin^2\theta } \\
\cos^2\Theta & = & \frac{(1-{\bar a}^2)\cos^2\theta}
{1- {\bar a}^2\cos^2\theta +{\bar a}^2w^2\sin^2\theta} \,
\end{eqnarray}
describe the coordinate transformation
$(w,\theta)\mapsto (\rho,\Theta)$.
 Moreover, we have
\[ \cos^2\theta  =  \frac{(1+{\bar a}^2w^2)\cos^2\Theta}
{1- {\bar a}^2\sin^2\Theta +{\bar a}^2w^2\cos^2\Theta} \, ,\]
\[ \sin^2\theta  =  \frac{(1-{\bar a}^2)\sin^2\Theta}
{1- {\bar a}^2\sin^2\Theta +{\bar a}^2w^2\cos^2\Theta} \, ,\]
\[ \rho^2 = \frac{1- {\bar a}^2\sin^2\Theta +{\bar a}^2w^2\cos^2\Theta}
{1+{\bar a}^2w^2}w^2  \, ,\]
\[ \phi= \frac{\Phi - {\bar a}{\bar t}}{1-{\bar a}^2} \, ,
 \quad  \rd\phi= \frac{\rd\Phi - {\bar a}\rd{\bar t}}{1-{\bar a}^2} \, ,\]
\[ w^2 =2\rho^2 \left[ 1- {\bar a}^2 \sin^2\Theta - {\bar a}^2\rho^2 +
\sqrt{(1- {\bar a}^2 \sin^2\Theta - {\bar a}^2\rho^2)^2
+4{\bar a}^2\rho^2\cos^2\Theta} \right]^{-1} \, ,\]

\[ \rd\theta= \frac{\sqrt{(1-{\bar a}^2)(1+{\bar a}^2w^2)}}
{1- {\bar a}^2\sin^2\Theta +{\bar a}^2w^2\cos^2\Theta}
\left(\rd\Theta-
\frac{{\bar a}^2 w\sin\Theta\cos\Theta}{1+{\bar a}^2 w^2}\rd w\right) \, . \]

\section{Curvature of conformal boundary in NUT-AdS spacetime}
For the metric (\ref{h0NUT}) we have the following non-vanishing
Christoffel symbols:
\begin{eqnarray}
\Gamma^0{_{\theta\phi}} & = & -{\bar l}\sin\theta(1+4{\bar l}^2)
\tan^2\frac\theta{2} \, , \\
\Gamma^0{_{\theta 0}} & = & 2{\bar l}^2\tan\frac\theta{2}\, , \\
\Gamma^\theta{_{\phi\phi}} & = & \sin^2\theta\left(
 4{\bar l}^2\tan\frac\theta{2} -\cot\theta\right) \, , \\
\Gamma^\theta{_{0\phi}} & = & - {\bar l}\sin\theta\, , \\
\Gamma^\phi{_{\theta 0}} & = & \frac{\bar l}{\sin\theta}\, , \\
\Gamma^\phi{_{\theta\phi}} & = & -2{\bar l}^2\tan\frac\theta{2}+\cot\theta \, .
\end{eqnarray}
For the Ricci tensor
\[ {\cal R}_{ab}=\partial_c \Gamma^c{_{ab}}- \partial_b \Gamma^c{_{ac}}
+ \Gamma^c{_{ab}}\Gamma^d{_{cd}} - \Gamma^d{_{ac}}\Gamma^c{_{db}} \]
we get the following components:
\begin{eqnarray}
{\cal R}_{\theta\theta} & = & 1+2{\bar l}^2 \, , \\
{\cal R}_{00} & = & 2{\bar l}^2 \, , \\
{\cal R}_{\phi\phi} & = & \sin^2\theta\left(1+
 2{\bar l}^2+8{\bar l}^4\tan^2\frac\theta{2}\right) \, , \\
{\cal R}_{0\phi} & = & - 4{\bar l}^3\sin\theta\tan\frac\theta{2}\, , \\
{\cal R}_{\theta 0} & = & 0\, , \\
{\cal R}_{\theta\phi} & = & 0 \, ,
\end{eqnarray}
together with the scalar curvature
\[ {\cal R} = h^{ab}{\cal R}_{ab}= 8(1+{\bar l}^2) \, . \]
The above formulae for the Ricci tensor imply
\[ {\cal R}_{0\theta |\phi}-{\cal R}_{0\phi |\theta} = 2 {\bar l}\sin\theta \, .
 \]
Moreover,
\[  \sqrt{-\det \stackrel{(0)}{h}} \beta^0{_0} = 2 C_{0\theta\phi}=
   2{\cal R}_{0\theta |\phi}-2{\cal R}_{0\phi |\theta} \]
because ${\cal R}_{,a}=0$. Similarly we have
\[ C_{0\theta 0}=0=C_{0\phi 0} \]
which implies
\[ \beta^\phi{_0}=0=\beta^\theta{_0} \, . \]


\begin{thebibliography}{66}
\bibitem{Abbott-Deser} L.F. Abbott and S. Deser,
\emph{Stability Of Gravity With A Cosmological Constant}, Nucl.
Phys. B 195, (1982) 76--96

\bibitem{genAdS}
{Anderson, Michael T. and Chru{\'s}ciel, Piotr T. and Delay, Erwann},
     \emph{Non-trivial, static, geodesically complete space-times with a
              negative cosmological constant. {II}. {$n\geq5$}},
 in {AdS/CFT correspondence: Einstein metrics and their conformal
              boundaries}, {IRMA Lect. Math. Theor. Phys.},
    vol. {8}, {165--204}, {Eur. Math. Soc., Z\"urich}     (2005);
     \emph{Non-trivial, static, geodesically complete, vacuum space-times
              with a negative cosmological constant},
   {J. High Energy Phys.} {10},
      (2002),  {063, 27};
Chrusciel, P. T. and Delay, E.,
     \emph{Non-singular, vacuum, stationary space-times with a
                  negative cosmological constant}, 2005,
                  \arxiv{gr-qc/0512110}
\bibitem {LarsAnd} L. Andersson, \emph{Bel--Robinson Energy and Constant
Mean Curvature Foliations}, Ann. Henri Poincar\'{e} 5, (2004)
235--244
\bibitem{AM} A. Ashtekar and A. Magnon,
\emph{Asymptotically anti-de Sitter spacetimes},
Classical and Quantum Gravity Lett. 1, (1984) L39
\bibitem{AASD} Abhay Ashtekar and Saurya Das,
\emph{Asymptotically Anti-de Sitter Space-times:
Conserved Quantities}, Class. Quantum Grav. 17, (2000) L17-L30,
\arxiv{hep-th/9911230}
\bibitem{BFGK} H. Baum, Th. Friedrich, R. Grunewald, I. Kath,
{\em Twistor and Killing Spinors on Riemannian
Manifolds}, Teubner-Verlag, Stuttgart-Leipzig (1991)
\bibitem{BCK} I.M. Benn, P. Charlton and J. Kress,
\emph{Debye potentials for Maxwell and Dirac fields from a
generalization of the Killing--Yano equation},
 Journal of Mathematical Physics 38, (1997) 4504--4527
\bibitem{Berg} G. Bergqvist, P. Lankinen, \emph{Unique characterization
 of the Bel--Robinson tensor}, Classical and Quantum Gravity 21,
 (2004) 3489--3503; G. Bergqvist, I. Eriksson and J.M.M. Senovilla,
 \emph{New electromagnetic conservation laws}, Classical and Quantum
 Gravity 20, (2003) 2663--68
\bibitem{Sen} M.A.G. Bonilla and J.M.M. Senovilla,
 \emph{Some Properties of the Bel and Bel--Robinson Tensors}
 General Relativity and Gravitation 29, (1997) 91--116
 \bibitem{Ch-Kl} D. Christodoulou and S. Klainerman, \emph{Asymptotic
Properties of Linear Field Equations in Minkowski Space},
Communications on Pure and Applied Mathematics 43, (1990) 137--199
\bibitem{CJK} P.T. Chru\'sciel, J. Jezierski and J. Kijowski,
\emph{Hamiltonian Field Theory in the Radiating Regime},
 Lecture Notes in Physics m 70, Springer 2002
 \bibitem{Douglas} S.R. Douglas, \emph{Letter: Review of the Definitions
of the Bel and Bel-Robinson Tensors}, General Relativity and
Gravitation 35, (2003) 1691--97
\bibitem{FG} C. Fefferman and C.R. Graham, \emph{Conformal Invariants},
Soc. Math. de France Ast\'erisque, hors s\'erie, (1985) 95--116
\bibitem{Hfriedrich} H. Friedrich,
\emph{Einstein equations and conformal structure:
Existence of Anti-de Sitter-type space-times},
    Journal of Geometry and Physics 17, (1995) 125--184.
\bibitem{HfriedrichC} H. Friedrich, \emph{Conformal Einstein evolution},
In: The Conformal Structure of Space-Times, Geometry, Analysis, Numerics
Series: Lecture Notes in Physics, Vol. 604,
J. Frauendiener, H. Friedrich (Eds.) 2002, \arxiv{gr-qc/0209018}


\bibitem{Glass-Naber} E.N. Glass and M.G. Naber, \emph{Gravitational mass
anomaly}, Journal of Mathematical Physics 35, (1994) 4178--83
\bibitem{Goldberg1} J.N. Goldberg,  \emph{Conserved quantities
at spatial and null infinity: The Penrose potential}, Phys. Rev. D
41, (1990) 410--417
\bibitem{Graham1} C.R. Graham, \emph{Volume and Area Renormalization for
Conformally Compact Einstein Metrics}, Rend. Circ. Mat. Palermo, suppl.,
63, (2000) 31–42, \arxiv{math.DG/9909042}


\bibitem{GP} J.B. Griffiths and J. Podolsky,
\emph{A new look at the Pleba\'nski--Demia\'nski family of solutions},
Int. J. Mod. Phys. D15, (2006) 335-370, \arxiv{gr-qc/0511091}
\bibitem{HT} M. Henneaux, C. Teitelboim, \emph{Asymptotically Anti-de Sitter Spaces},
Commun. Math. Phys. 98, (1985) 391--424
\bibitem{HIM} Stefan Hollands, Akihiro Ishibashi, Donald Marolf,
\emph{Comparison between various notions of conserved charges in asymptotically
AdS-spacetimes}, Classical and Quantum Gravity 22, (2005)
2881--2920, \arxiv{hep-th/0503045}
\bibitem{HM} G.T. Horowitz and R.C. Myers, \emph{AdS correspondence
and a new positive energy conjecture for general relativity},
Phys. Rev. D 59, (1998) 026005
\bibitem{InfeldSchild} L. Infeld and A. Schild, \emph{A New Approach to
Kinematic Cosmology}, Phys. Rev. 68 (1945) 250-272
\bibitem{JJML} J. Jezierski, M. {\L}ukasik,
\emph{Conformal Yano-Killing tensor for the Kerr metric
and conserved quantities},
Classical and Quantum Gravity {23} (2006) 2895--2918, \arxiv{gr-qc/0510058}
\bibitem{JJMLnut} J. Jezierski, M. {\L}ukasik,
\emph{Conformal Yano-Killing tensors for the Taub-NUT metric},
Classical and Quantum Gravity {24} (2007) 1331--1340,
\arxiv{gr-qc/0610090}









\bibitem{JJspin2} J. Jezierski, \emph{The Relation between Metric and
Spin--2 Formulations of Linearized Einstein Theory},
 Gen. Rel. and Grav. 27, (1995) 821--843, \arxiv{gr-qc/9411066}
\bibitem{kerrnut} J. Jezierski,
 {\it  Conformal Yano--Killing tensors and asymptotic CYK tensors for
 the Schwarzschild metric},
 Classical and Quantum Gravity 14, (1997) 1679--1688, \arxiv{hep-th/9411074}
\bibitem{cykem} J. Jezierski, {\it CYK Tensors, Maxwell Field
   and Conserved Quantities for Spin-2 Field},
   Classical and Quantum Gravity {19}, (2002) 4405--4429, \arxiv{gr-qc/0211039}





\bibitem{GRGKij} J. Kijowski, \emph{A Simple Derivation of Canonical Structure and
Quasi-local Hamiltonians in General Relativity},
General Relativity and Gravitation, Vol. 29, No. 3, (1997) 307--343
\bibitem{Moroianu} A. Moroianu, U. Semmelmann,
\emph{Twistor forms on K\"ahler manifolds}, Ann. Scuola Norm. Sup.
Pisa Cl. Sci. (4) II (2003), 823--845, \arxiv{math.DG/0204322}






\bibitem{Pen1} R. Penrose, \emph{Quasi-local Mass and Angular Momentum in
General Relativity}, Proc. Roy. Soc. Lond. A381, (1982) 53--62
\bibitem{Pen-Rin} R. Penrose and W. Rindler, {\it Spinors and Space-time},
Cambridge University Press, Vol. 2, p. 396 (Cambridge 1986)

\bibitem{Rindler} W. Rindler, \emph{Relativity: special, general, and cosmological},
Oxford Univ. Press (2001)
\bibitem{Skenderis} K. Skenderis, \emph{Asymptotically Anti-de Sitter Spacetimes
and their stress energy tensor}, Int. J. Mod. Phys. A16, (2001) 740-749,
 \arxiv{hep-th/0010138}
\bibitem{Skenderis2} Miranda C.N. Cheng and Kostas Skenderis,
 \emph{Positivity of energy for asymptotically locally AdS
spacetimes},  J. High Energy Phys. JHEP08, (2005) 107,
   \arxiv{hep-th/0506123}
\bibitem{Semmelmann} U. Semmelmann, \emph{Conformal Killing forms on
Riemannian manifolds}, Mathematische Zeitschrift 245, (2003)
503--527

\bibitem{Stepanow} S.E. Stepanov, \emph{The Vector Space of Conformal
 Killing Forms on a Riemannian Manifold},
 Journal of Mathematical Sciences 110, (2002) 2892--2906;
\emph{On conformal Killing 2-form of the electromagnetic field},
Journal of Geometry and Physics 33, 
(2000) 191--209

\bibitem{SKM} H. Stephani et al., \emph{Exact solutions of Einstein's
 field equations}, 2nd ed., University Press (Cambridge 2003)


\bibitem{Tachibana} S. Tachibana,
\emph{On conformal Killing tensor in a Riemannian space}, Tohoku
Math. J. (2) 21, (1969) 56--64; T. Kashiwada,  \emph{On conformal
Killing tensor}, Natur. Sci. Rep. Ochanomizu Univ. 19, (1968)
67--74; S. Tachibana  and T. Kashiwada,  \emph{On the
integrability of Killing--Yano's equation}, J. Math. Soc. Japan
21, (1969) 259--265


\bibitem{Yano} K. Yano, \emph{Some remarks on tensor fields and curvature},
Ann. Math. {55}, (1952) 328--347

\end{thebibliography}
\end{document}